\documentclass[11pt]{article}
\usepackage{geometry,cite,verbatim}                
\usepackage[utf8]{inputenc}
\usepackage{amsmath, amsthm, amsfonts, amssymb,mathrsfs}
\usepackage{graphicx}
\usepackage{lmodern}
\usepackage{mathabx}
\usepackage{color,xcolor}
\usepackage{esint,physics}
\usepackage{caption,subcaption,verbatim}
\usepackage{wrapfig}
\usepackage{hyperref}
\usepackage{multicol}
\usepackage[bottom]{footmisc}

\newtheorem{theorem}{Theorem}[section]

\newtheorem{proposition}[theorem]{Proposition}

\newtheorem{remark}[theorem]{Remark}

\author{Paul Cazeaux
\thanks{Department of Mathematics, Virginia Tech, Blacksburg, VA 24060, \href{mailto:cazeaux@vt.edu}{\texttt{cazeaux@vt.edu}}. ORCID: 0000-0003-2972-2616. PC's research was supported in part by National Science Foundation Award DMS-189220 and Simons Collaboration Grants for Mathematicians No. 966604.},
Drake Clark
\thanks{School of Mathematics, University of Minnesota, Minneapolis, MN 55408,
\href{mailto:clar1939@umn.edu}{\texttt{clar1939@umn.edu}}. DC's research was supported in part by National Science Foundation Award DMS-1906129.},
Rebecca Engelke,
\thanks{Department of Physics, Harvard University, Cambridge, Massachusetts 02138, \href{mailto:rebeccaengelke@gmail.com}{\texttt{rebeccaengelke@gmail.com}}, \href{mailto:philipkim@g.harvard.edu}{\texttt{philipkim@g.harvard.edu}}. RE's and PK's research was supported in part by National Science Foundation Award DMREF Award No. 1922165.},
Philip Kim\footnotemark[3],
Mitchell Luskin
\thanks{School of Mathematics, University of Minnesota, Minneapolis, MN 55408,
\href{mailto:luskin@umn.edu}{\texttt{luskin@umn.edu} (Corresponding author)}. ORCID: 0000-0003-1981-199X. ML's research was supported in part by National Science Foundation Award DMREF Award No. 1922165 and Simons Targeted Grant Award No. 896630.  }
}
\newcommand{\R}{\mathbb{R}}

\newcommand{\Z}{\mathbb{Z}}

\newcommand{\La}{\mathcal{R}}

\newcommand{\bb}{\mathbf{b}}
\newcommand{\be}{\mathbf{e}}
\newcommand{\bu}{\mathbf{u}}
\newcommand{\bv}{\mathbf{v}}
\newcommand{\bx}{\mathbf{x}}
\newcommand{\bR}{\mathbf{R}}
\newcommand{\ba}{\mathbf{a}}
\newcommand{\bgamma}{{\boldsymbol\gamma}}
\newcommand{\bxi}{{\boldsymbol\xi}}
\newcommand{\bk}{\mathbf{k}}

\newcommand{\nmod}{\text{mod}}

\newcommand{\pc}[1]{{{\color{black} #1}}}

\newcommand{\cpc}[1]{{{\color{black}{[PC: #1]}}}}
\newcommand{\ml}[1]{{\color{black} #1}}

\newcommand{\dc}[1]{{\color{black} #1}}

\newcommand{\bGamma}{\Gamma}
\title{Relaxation and domain wall structure of bilayer moir\'e systems}

\begin{document}
\maketitle
\begin{abstract}
Moir\'e patterns result from setting a 2D material such as graphene on another 2D material with a small twist angle or from
the lattice mismatch of 2D heterostructures.  We present a continuum model for the elastic energy of these bilayer moir\'e structures that includes an intralayer elastic energy and an interlayer misfit energy
that is minimized at two stackings (disregistries).
We show by theory and computation that the displacement field that minimizes the global elastic energy subject to a global boundary constraint
gives large alternating regions of one of the two energy-minimizing stackings separated by domain walls.

We derive a model for the domain wall structure from the continuum bilayer energy and give a rigorous asymptotic estimate
for the structure.  We also give an improved estimate for the $L^2$-norm of the gradient on the moir\'e unit cell for twisted bilayers that scales at most \textit{inversely linearly} with the twist angle, a result which is consistent with the formation of one-dimensional domain walls with a fixed width around triangular domains at very small twist angles.
\end{abstract}
\section{Introduction.}

\begin{figure}[t!]
    \centering
    \includegraphics[width=.9\textwidth]{"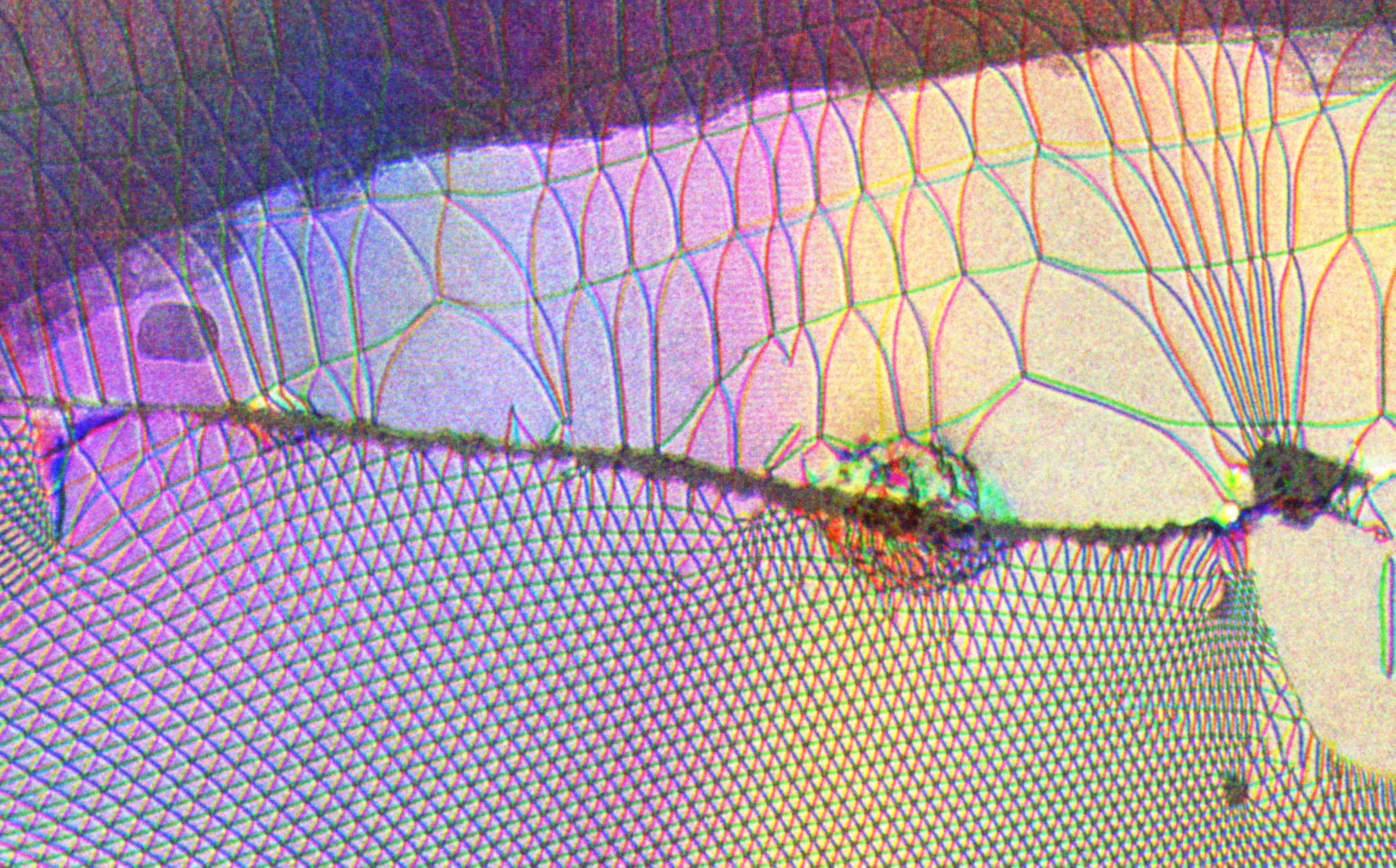"}
    \caption{Composite dark field image of the moiré domains in a MoSe$_2$/WSe$_2$ heterostructure. A horizontal crack separates two regions, with the bottom region exhibiting a mostly uniform twist angle of 1$^\circ$, and the top region having a non-uniform strain dominated by a .6$^\circ$ twist with large shear in the upper right region. Domain walls are imaged in the dark field transmission electron microscope using the second order Bragg peaks and superposed together to provide contrast according to the three different Burgers vectors (see Fig.~\ref{fig:figure5}), colored red, green, and blue, respectively.
    }
    \label{fig:figure1}
\end{figure}
Recent scientific and technological interest in van der Waals 2D materials such as twisted bilayer graphene or transition metal dichalcogenides (TMDs) such as molybdenum disulfide ($\text{MoS}_2$) has motivated the investigation of the
complex microstructure of bilayer moir\'e materials (see Figure \ref{fig:figure1}).  The energy landscape with respect to the possible bilayer stackings is not uniform \cite{alden2013strain}
(see Figures \ref{fig:figure4}, \ref{fig:figure6a}, and \ref{fig:figure6b}), so the twisted bilayer structure mechanically relaxes to enhance the regions of lowest stacking energy within each moir\'e cell while confining the regions of highest stacking energy to domain walls separating the moir\'e cells so as to maintain a global twist angle constraint \cite{alden2013strain,KimRelax18} (see Fig.~\ref{fig:figure9}).

Models for the relaxation of the structure of twisted bilayer materials have been proposed in \cite{dai2016twisted,relaxphysics18}. A general and rigorous model for the relaxation of incommensurate multilayer materials was proposed in \cite{cazeaux2020energy} that also includes untwisted heterostructures which have moir\'e superlattices from lattice mismatch (such as untwisted $\text{WSe}_2/\text{MoS}_2$) and trilayer systems which have more complex moir\'e of moir\'e structure \cite{zhu2019moir}.

Transferring a two-dimensional lattice onto another with a small rotation $\theta$ gives rise to a moir\'e superlattice with lattice constant $a_M=a_0/(2\sin(\theta/2))$ that is much larger than the single layer lattice constant $a_0$ (see Figure .~\ref{fig:figure8}a). The capability to fabricate twisted 2D bilayers such as graphene or $\text{MoS}_2$ with precise control of small twist angles as low as $\theta\approx .1^\circ$ or smaller has opened a new era in the development and design of materials with previously inaccessible electronic, magnetic, and optical properties.  For example, the discovery of Mott insulator and superconducting electronic phases in twisted bilayer graphene at the ``magic angle'' of $\theta=1.05^\circ$ \cite{Cao2018a,Cao2018} has led to a new era in the investigation of strongly correlated electronic and magnetic phases.

Low energy, continuum models for the electronic density of states, band structure, and transport properties of twisted bilayer materials
\cite{Bistritzer2011,rigorousbm22} were derived that predicted the existence of correlated physics (superconducting and Mott insulator phases) at the ``magic angle."
More general formulations using the local configuration or disregistry for incommensurate atomistic structures (without a supercell approximation) were later developed and rigorously analyzed \cite{dos17,genkubo17,kubocomp20,momentumspace17,kubocomp20}. All of these approaches assumed that each lattice remains rigid when twisted and then placed on another lattice, and hence each possible stacking (see Fig.~\ref{fig:figure3}) is sampled uniformly over the moir\'e unit cell.

The relaxation of bilayer structures can have a significant effect on the electronic properties such as band gap that
are predicted by the above models \cite{KimRelax18,relaxed_elect22}.
Recently developed models for the relaxation of twisted bilayer homostructures and heterostructures have been used to obtain more accurate computations for fundamental electronic properties
such as band gaps \cite{KimRelax18}.  A rigorous analysis of the effect of relaxation on the electronic density of states and other electronic properties and
a more efficient computational method has been recently given in
\cite{relaxed_elect22}.

This paper first reviews in Section \ref{subsec:disregistry} the concept of disregistry of one layer with respect to another layer
following the approach given in \cite{cazeaux2020energy,relaxphysics18} and is extended to the disregistry of relaxed modulated layers in Section \ref{subsec:model} following \cite{cazeaux2020energy}.  The concept of the generalized stacking fault energy is described in Section \ref{sec:gsfe}. The fundamental bilayer model developed in \cite{cazeaux2020energy,relaxphysics18} is then given in Section \ref{subsec:model}.

We derive a model for the domain wall structure from the bilayer energy \eqref{eq:genPN} in Section \ref{sec:dw} and give a rigorous asymptotic estimate
in Proposition \ref{prop:dw} for the structure.
We also present detailed computational results for the domain wall structure and compare to our theoretical results.

In Section \ref{sec:4types}, we derive the moir\'e domain orientation and lengths scale and present computational results for each of 4 ``pure" strain fields: \textit{twisted}, isotropic strain, pure shear, and
\ml{simple shear.}
We note that each of these moir\'e orientations can be found in non-uniformly strained bilayer structures (see Figure \ref{fig:figure1} and \cite{nonabelian}).
For twisted bilayers, we also give an improved estimate for the $L^2$-norm of the gradient on the moir\'e unit cell (Theorem ~\ref{thm:est}) that scales at most \textit{inversely linearly} with the twist angle which is consistent with the formation of one-dimensional domain walls with a fixed width around triangular domains at very small twist angles, as observed in simulations (see~Figure \ref{fig:figure9}) and experiments~\cite{KimRelax18}.

\section{Continuum model for mismatched bilayers}
\subsection{Geometry}
We consider stacks of two-dimensional crystalline layers, where in the reference state atoms
\begin{wrapfigure}{r}{.4\textwidth}
  \centering
  \includegraphics[width=.4\textwidth]{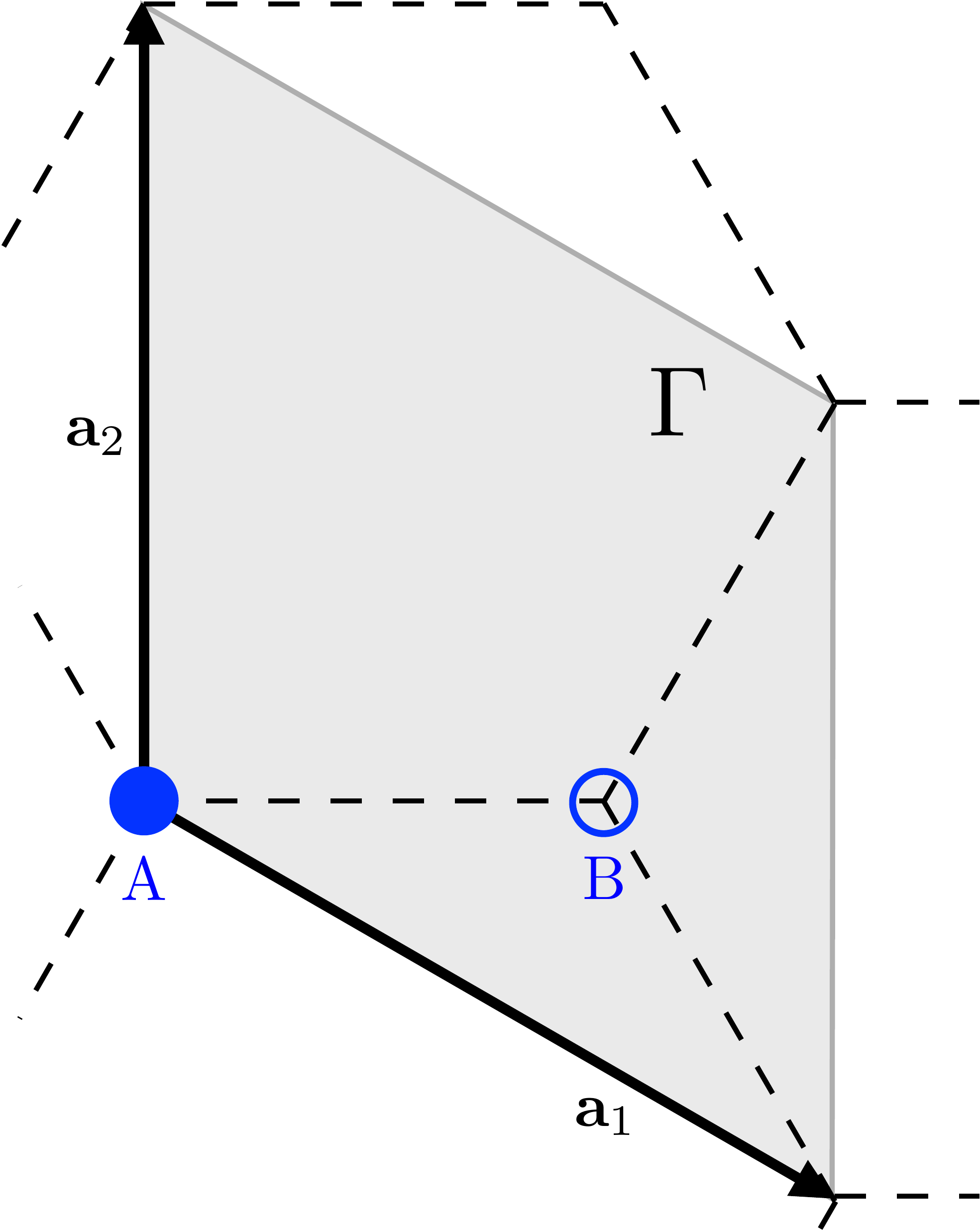}
  \caption{\ml{Unit cell $\Gamma$ (gray) of graphene given by lattice vectors
   $\ba_1$ and $\ba_2$ with sublattices A and B.}}\label{fig:figure2}
\end{wrapfigure}
are distributed periodically in each layer according to a Bravais lattice.
For example, the triangular lattice associated with graphene can be described by the fundamental matrix whose columns are generating lattice vectors:
\begin{equation}\label{def:GrapheneLattice}
  A =  a_0 \begin{pmatrix}
    \sqrt{3}/2 & 0 \\ -1/2 & 1
  \end{pmatrix},
\end{equation}
with $a_0$ the lattice constant, such that the lattice sites $\La$ are given as $A \begin{pmatrix}
  m \\ n
\end{pmatrix}$
with $m,n$ integers.
The unit cell of the lattice is given by
\[
  \Gamma := \left \{ A \begin{pmatrix}
    s \\ t
  \end{pmatrix}  \quad \text{ such that } 0 \leq s,t < 1\right \},
\]
which is equipped with periodic boundary conditions and can also be seen as the torus $\R^2 / A \Z^2$.
Note that there may be more than one atom in each unit cell: for example, graphene has two atoms (typically denoted $A$ and $B$) in each unit cell (see Figure \eqref{fig:figure2}), while molybdenum disulfide ($\text{MoS}_2$) has three.

We are particularly interested in stackings where a moir\'e pattern emerges, i.e., where the layers, being almost aligned, have \textit{almost} but \textit{not exactly} the same fundamental matrix. One particular example is the famous \textit{twisted bilayer} of graphene which exhibits superconductivity at the small so-called magic twist angle $\theta = 1.1^\circ$: with our notation, this is obtained by rotating slightly each layer in opposite directions, with the fundamental matrices for each layer being given by:
\[
  A_1 = R_{-\theta/2} A, \qquad A_2 = R_{\theta/2} A, \qquad \text{with } R_\phi = \begin{pmatrix}
    \cos \phi & -\sin \phi \\ \sin \phi & \cos \phi
  \end{pmatrix}.
\]
More generally, moir\'e patterns in bilayers will be obtained whenever the fundamental matrices of the lattices satisfy the condition
\begin{equation}\label{eq:MoireCondition}
  \Vert A_1 A_2^{-1} - I \Vert, \  \Vert A_2 A_1^{-1} - I \Vert \ll 1,
\end{equation}
where $I$ stands for the identity matrix.
\subsection{Disregistry}\label{subsec:disregistry}
\begin{figure}[b!]
  \centering
  \subcaptionbox{Twisted bilayer moir\'e geometry\label{fig:figure3a}}[.9\textwidth]
  {\includegraphics[width=.7\textwidth]{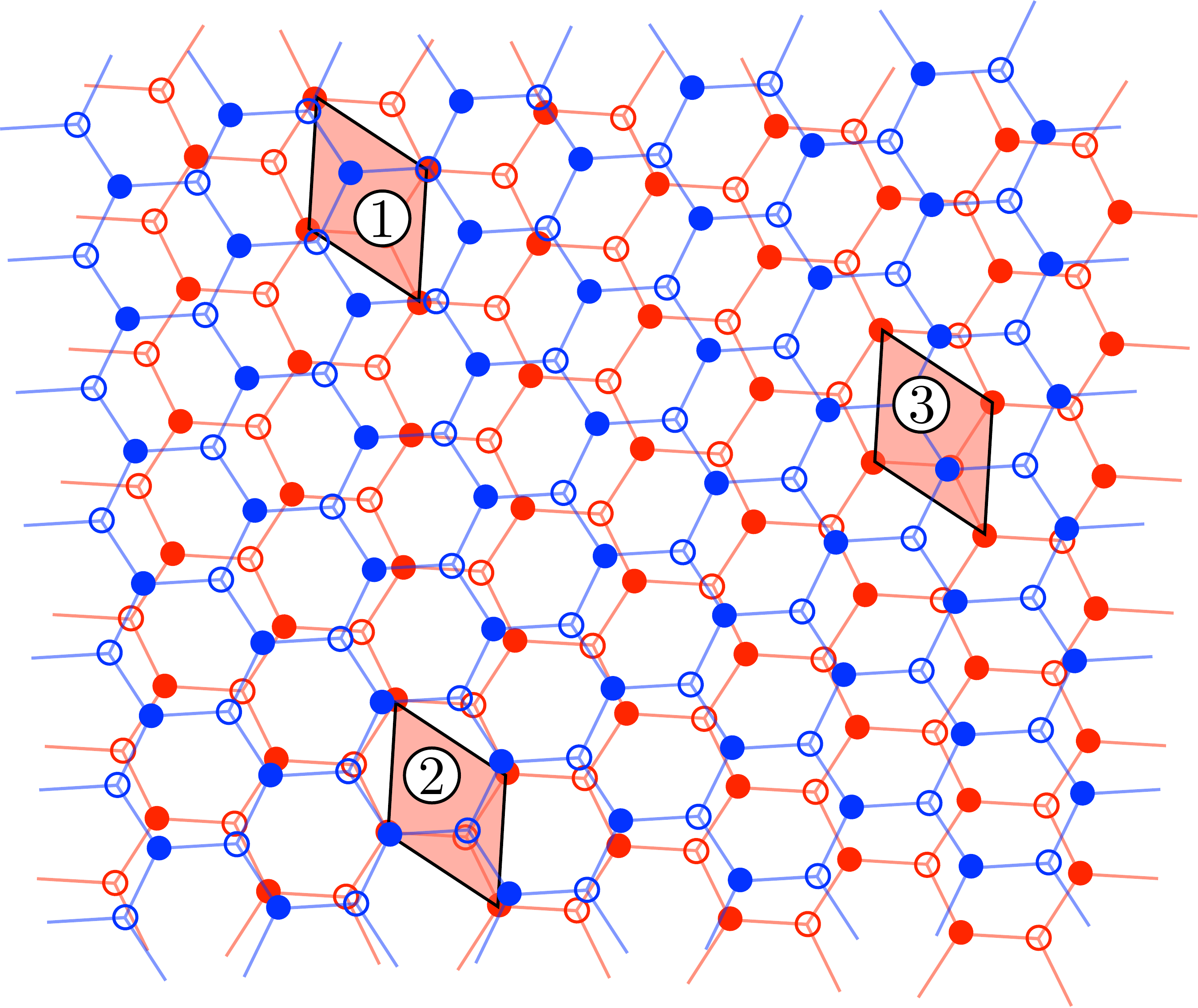}}
  \medskip

  \subcaptionbox{AB configuration\label{fig:figure3b}}[.29\textwidth]
  {\includegraphics[height=1.5in]{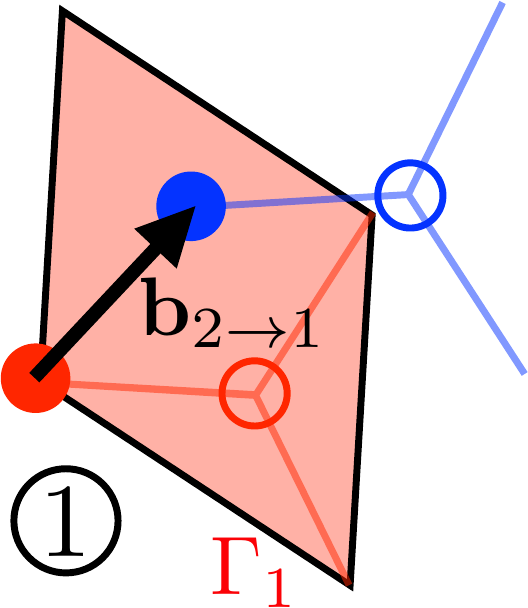}}
  \subcaptionbox{AA configuration: $\mathbf{b}_{2 \to 1} \approx \boldsymbol{0}$\label{fig:figure3d}}[.4\textwidth]
  {\includegraphics[height=1.5in]{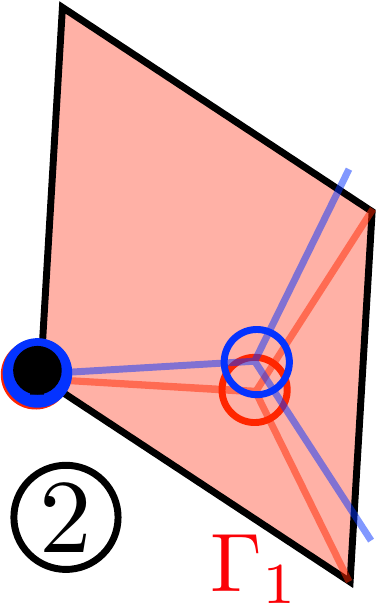}}
  \subcaptionbox{BA configuration\label{fig:figure3c}}[.29\textwidth]
  {\includegraphics[height=1.5in]{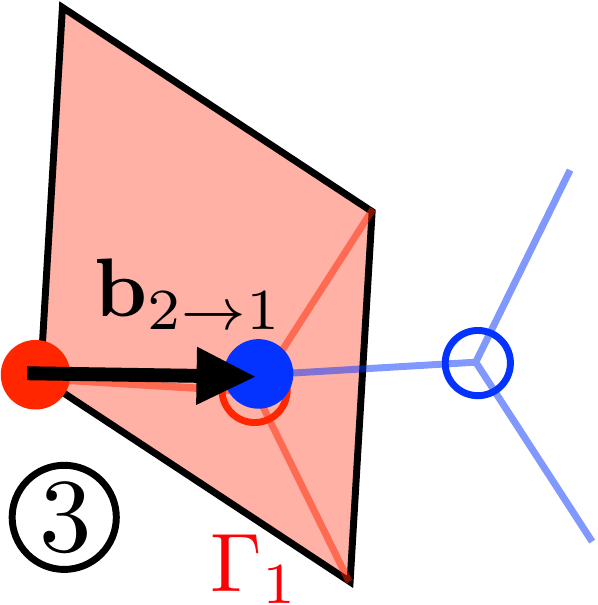}}

  \caption{Configurations and disregistry for identical hexagonal lattices with the large twist angle $\theta = 6.8^\circ$. Three special configurations are identified within the moir\'e lattice, with close-ups depicting the disregistry $\mathbf{b}_{2 \to 1}$ of layer 2 (in blue) with respect to layer 1 (in red) in each case.}
  \label{fig:figure3}
\end{figure}


Following~\cite{relaxphysics18}, we assume without loss of generality that the disregistry is zero at the origin, and we define the disregistry of $\bR_1\in\La_1$ with
respect to $\La_2$ and $\bR_2\in\La_2$ with
respect to $\La_1$ of the unrelaxed bilayer structure by (see Figure~\ref{fig:figure3}):
\begin{equation*}
\bb_{1\to 2}(\bR_1) = \nmod_{\Gamma_2}(\bR_1)\quad\text{and}\quad \bb_{2\to 1}(\bR_2) = \nmod_{\Gamma_1}(\bR_2),
\end{equation*}
where $\nmod_{\Gamma_2}(\bR_1) = \bR_1+\widetilde{\bR_2} \in \Gamma_2$ for appropriate $\widetilde{\bR_2} \in \La_2.$
We have derived interpolation formulae for undistorted lattices \cite{relaxphysics18,cazeaux2020energy}
\begin{equation}\label{eq:disregistry}
  \bb_{1\to2}(\bx) := (I - A_2A_1^{-1}) \bx \quad (\textrm{mod } \bGamma_2)\quad\text{and}\quad \bb_{2\to1}(\bx) := (I - A_1 A_2^{-1}) \bx \quad (\textrm{mod } \bGamma_1),
\end{equation}
which define slowly varying disregistries in space due to assumption~\eqref{eq:MoireCondition}, although the disregistry is a fundamentally discrete / atomistic quantity.
\noindent It is easy to check that, provided~\eqref{eq:MoireCondition} holds:
\[
    \bb_{1\to2}(\bx)=-A_2A_1^{-1}\bb_{2\to1}(\bx)\quad (\textrm{mod } \bGamma_2),\quad \text{so}\ \bb_{1\to2}(\bx)\approx-\bb_{2\to1}(\bx).
\]
\begin{remark}
  Note that we follow here the disregistry convention from~\cite{relaxphysics18}, which is opposite to~\cite{cazeaux2020energy}.
\end{remark}
We note that both disregistries are periodic as a function of the continuous position variable $\bx$, leading us to identify the lattice and unit cell of the bilayer moir\'e structure
(see Figure~\ref{fig:figure8}),
\begin{equation}\label{def:moirelattice}
  A_\mathcal{M} := (A_1^{-1} - A_2^{-1})^{-1}, \qquad \Gamma_\mathcal{M} := A_\mathcal{M} [0,1)^2,
\end{equation}
since $\bb_{1\to 2}(\bx)$ and $\bb_{2\to 1}(\bx)$ are isomorphisms \cite{relaxphysics18,cazeaux2020energy}
\begin{equation*}\label{def:MoireMappings}\begin{split}
    &\bb_{1\to 2}: \left \{ \begin{aligned}
        \Gamma_\mathcal{M} &\to \Gamma_2
        \\ \bx &\mapsto (I-A_2A^{-1}_1) \bx =(I-A_2A^{-1}_1) \bx+A_2(\be_1+\be_2)\quad (\textrm{mod } \bGamma_2),
    \end{aligned} \right. \\
    &\bb_{2\to 1}: \left \{ \begin{aligned}
        \Gamma_\mathcal{M} &\to \Gamma_1, \\ \bx &\mapsto (I-A_1A^{-1}_2) \bx \quad (\textrm{mod } \bGamma_1).
    \end{aligned} \right.
    \end{split}
\end{equation*}
\begin{wrapfigure}{r}{.35\textwidth}
    \centering
    \includegraphics[width=.35\textwidth]{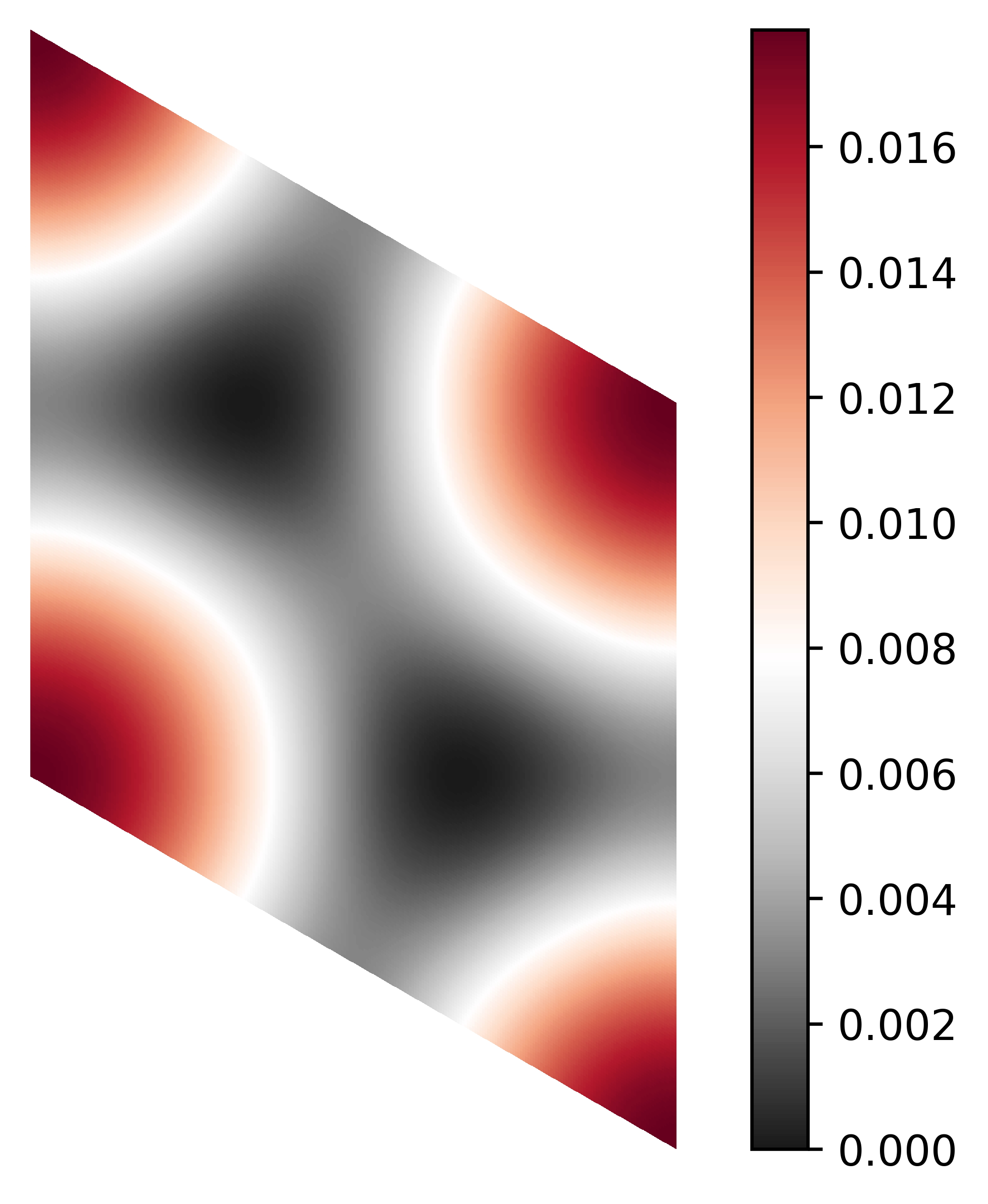}
    \caption{GSFE approximation for bilayer graphene as a function of disregistry from \cite{srolovitzgsfe} plotted over the unit cell and shifted so that the minimum occurs at 0. }
    \label{fig:figure4}
    \vspace{-.6in}
\end{wrapfigure}

\ml{\begin{remark}
    We note that $\bx \mapsto (I-A_2A^{-1}_1) \bx$ maps
    $\Gamma_\mathcal{M} \to -\Gamma_2,$ so we can add $A_2(\be_1+\be_2)$ to obtain
    that
    $ \bx \mapsto(I-A_2A^{-1}_1) \bx+A_2(\be_1+\be_2)$ maps $\Gamma_\mathcal{M} \to \Gamma_2.$
\end{remark}}

\begin{remark}
    We note that the mappings $\bb_{1\to 2}$ and $\bb_{2\to 1}$ are distinct group homomorphisms of the fundamental group of the torus.
    When the layers are respectively twisted by an angle $\theta>0$,
    $A_1 = R_{-\theta/2} A$ and $A_2 = R_{+\theta/2} A,
$
 and  we can verify that  $\curl \bb_{1\to 2}=-2\sin \theta<0 $ and $\curl \bb_{2\to 1}=2\sin \theta>0 $ .
\end{remark}
\noindent In most cases, the underlying layer lattices are incommensurate, so the moir\'e lattice is only an approximate periodicity of the system. However, we will see that in the continuum approximation the bilayer system is exactly periodic at the moiré scale; however, this is not the case for systems composed of three or more layers \cite{zhu2019moir}.

\subsection{Generalized stacking fault energy }\label{sec:gsfe}

The generalized stacking fault energy (GSFE) accounts for local contributions to the global
stacking energy and is assumed to depend only on the local configuration as measured by the disregistries.
The GFSE measures the energy cost of the layers not being uniformly in an optimal stacking configuration, but rather in a slowly varying disregistry induced by a global rotation or deformation of one of both layers.
As shown in Figure \ref{fig:figure4} for bilayer graphene, it is thus minimized at the AB and BA configurations and maximized at the AA configuration displayed in Figure~\ref{fig:figure3}. The GSFE is computed for aligned layers and is a good approximation for small rotations \cite{cazeaux2020energy}.

The generalized stacking fault energy for layer 1 with respect to layer 2 as a function of disregistry is given by $\Phi_1:\Gamma_2\to \R$   and the GFSE for layer 2 with respect to layer 1 is given by
$\Phi_2:\Gamma_1\to \R$ \cite{cazeaux2020energy}.  The GFSEs for unrelaxed bilayer structures at $\bx\in\R^2$ are thus given by
$\Phi_1 \left ( \bb_{1 \to 2}(\bx)  \right ) $ and $\Phi_2 \left ( \bb_{2 \to 1}(\bx)  \right )$.


\subsection{Bilayer model in real space}\label{subsec:model}
Our starting point will be the model developed in~\cite{cazeaux2020energy,relaxphysics18,dai2016twisted}, where the displacement minimizes an elastic energy that decomposes as:
\begin{equation}\label{eq:genPN}
  \mathcal{E}[\bu_1, \bu_2] = \sum_{j=1}^2 \mathcal{E}^j_\mathrm{intra}(\bu_j) + \mathcal{E}_\mathrm{inter}(\bu_1 - \bu_2),
\end{equation}
where we assume that the displacements $\bu_1(\bx)$ and $\bu_2(\bx)$ are periodic functions with respect to the moir\'e
lattice \eqref{def:moirelattice}.  This can be motivated by equivalently assuming that the displacements $\bu_1(\bx)$ and $\bu_2(\bx)$ are functions of the disregistries
$\bb_{1\to 2}(\bx)$ and $\bb_{2\to 1}(\bx),$ respectively \cite{cazeaux2020energy}.

The contribution $\mathcal{E}^j_\mathrm{intra}$ corresponds to the elastic cost of straining the layers (which we model here by isotropic linear elasticity):
\begin{equation}
  \begin{aligned}
    \mathcal{E}^1_\mathrm{intra}[\bu_1] &= \int_{\Gamma_\mathcal{M}} \frac{\lambda_1}{2} \,(\mathrm{div}\ \bu_1 )^2 + \mu_1\, \varepsilon(\bu_1) : \varepsilon(\bu_1) d \bx, \\
    \mathcal{E}^2_\mathrm{intra}[\bu_2] &= \int_{\Gamma_\mathcal{M}} \frac{\lambda_2}{2} \,(\mathrm{div}\ \bu_2 )^2 + \mu_2\, \varepsilon(\bu_2) : \varepsilon(\bu_2) d \bx,
  \end{aligned}
\end{equation}
where $\lambda_i$ and $\mu_i$ are Lam\'e parameters and $\varepsilon(\bu_i)$ are the linear strains for layer $i,$ while the contribution $\mathcal{E}_\mathrm{inter}$ corresponds to the misfit energy due to lattice misalignment between the two layers:
\begin{equation}\label{eq:misfitEnergy}
    \mathcal{E}_\mathrm{inter}[\bv] = \frac{1}{2} \int_{\Gamma_\mathcal{M}} \Phi_1 \left ( \bb_{1 \to 2}(\bx) + \bv(\bx) \right ) +  \Phi_2 \left ( \bb_{2 \to 1}(\bx) - \bv(\bx) \right ) d \bx,
\end{equation}
where \ml{$\bv = \bu_1 - \bu_2.$}
We note that the arguments of the generalized stacking fault energies $\Phi_1:\Gamma_2\to \R$ and $\Phi_2:\Gamma_1\to \R$
in $\mathcal{E}_\mathrm{inter}[\bv]$ are the modulated disregistries for the relative distortion of the lattice 
\begin{equation}
\widetilde{\bb}_{1 \to 2}(\bx):=\bb_{1 \to 2}(\bx) + \bv(\bx) \quad \text{and}\quad
\widetilde{\bb}_{2 \to 1}(\bx):=\bb_{2 \to 1}(\bx) - \bv(\bx).
\end{equation}
We look for a \textit{local} minimum of the energy (note that the global energy minimum would have both layers aligned in the optimal stacking configuration, but we are enforcing a fixed global misalignment through the periodic boundary conditions on the moiré torus).

\subsection{Interlayer symmetry}

A common thread will be that for two vertically stacked, possibly strained but initially identical layers and neglecting non-linear elastic effects, we can model the intralayer
elastic energies by the same values of Lamé parameters $\lambda$, $\mu$ between the two layers.
A natural symmetry then emerges: notice that $\mathcal{E}[\bu_1, \bu_2] = \mathcal{E}[-\bu_2, -\bu_1]$ because the misfit energy depends only on the difference $\bv = \bu_1 - \bu_2 = (- \bu_2) - (-\bu_1)$. In fact, the displacements are uniquely determined by the nonlinear forcing term in the linear elastic regime~\cite{cazeaux2020energy} as the Euler-Lagrange equations associated with the minimum read:
\[
    \left \{ \begin{aligned}
        - \mathbf{div} \left ( \lambda \, ( \mathrm{div}\ \bu_1) I + 2 \mu\,  \varepsilon(\bu_1) \right ) &= \mathbf{F}_\mathcal{M}(\bv), \\
        - \mathbf{div} \left ( \lambda \, ( \mathrm{div}\ \bu_2) I + 2 \mu\,  \varepsilon(\bu_2) \right ) &= -\mathbf{F}_\mathcal{M}(\bv),
    \end{aligned} \right.
\]
where \ml{$\mathbf{F}_\mathcal{M}[\bv](\bx) := \frac{1}{2} \nabla \Phi_1 \left ( \bb_{1 \to 2}(\bx)+  \bv(\bx) \right ) - \frac{1}{2} \nabla \Phi_2 \left ( \bb_{2 \to 1}(\bx) - \bv(\bx) \right )$.}
Uniqueness of the solution of the linear elasticity problem then implies:
\begin{equation}\label{eq:interlayersym}
    \bu := \bu_1 = - \bu_2,
\end{equation}
hence knowledge of the single planar displacement $\bu$ is sufficient to describe the displacement of both layers.
Furthermore, the total energy of the bilayer system thus simplifies to
\begin{equation}\label{eq:genPN_sym}
        \mathcal{E}[\bu] = 2 \mathcal{E}_\mathrm{intra}[\bu] + \mathcal{E}_\mathrm{inter}[2\bu],
\end{equation}
where $\displaystyle  \mathcal{E}_\mathrm{intra}[\bu] := \int_{\Gamma_\mathcal{M}} \frac{\lambda}{2} \, (\mathrm{div}\ \bu )^2 + \mu \, \varepsilon(\bu) : \varepsilon(\bu) d \bx$.

\section{Domain wall structure}\label{sec:dw}
Our first example will be one without a moiré structure --- but rather a single domain wall at a straight interface $y = 0$ within a bilayer structure in disregistry with a global rotation at angle $\phi$, with two half-planes $\pm y > 0$ both close to aligned, energy minimizing, non equivalent configurations, attained at $\pm y \to \infty$.
Such structures may be described as partial interlayer dislocations~\cite{dai2016structure} and modeled using the same generalized Peierls-Nabarro energy adapted from above~\eqref{eq:genPN}.

\subsection{Geometry}
More precisely, we assume here that the lattice fundamental matrices and initial disregistry are given by
\[
  A_1 = A_2 = R_\phi A, \quad \text{so} \quad \bb_{1\to2}(y) = -\bb_{2\to1}(y) = R_\phi \left [ \bb_\textrm{SP} +
 \tanh(\eta y)\Delta \bb \right ],
  \]
  where $\eta \to 0^+$ is a small parameter.

\begin{figure}[ht]
  \centering
  \subcaptionbox{\label{fig:figure5a}}[.3\textwidth]
  {\includegraphics[height=2in]{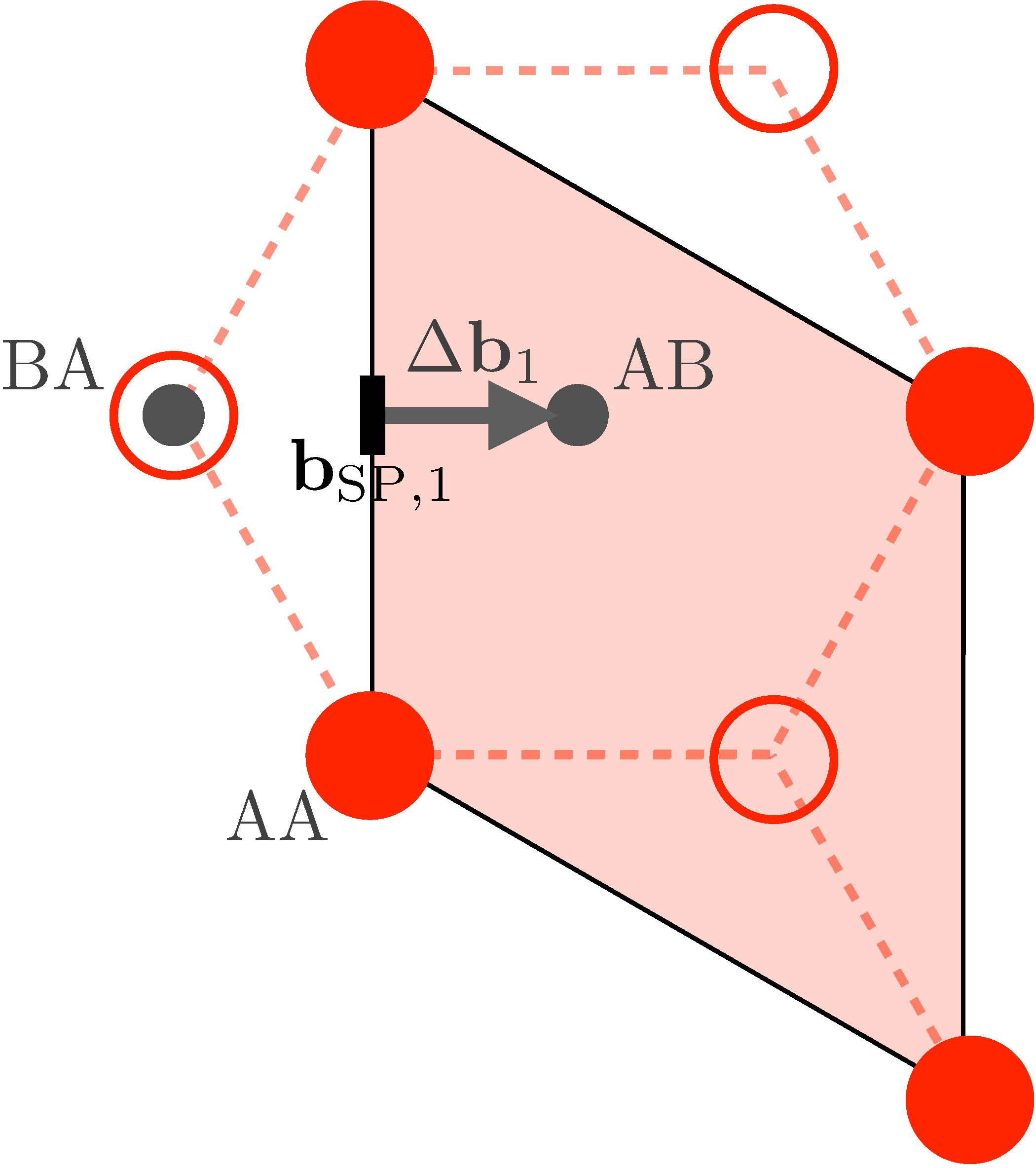}}
  \subcaptionbox{\label{fig:figure5b}}[.3\textwidth]
  {\includegraphics[height=2in]{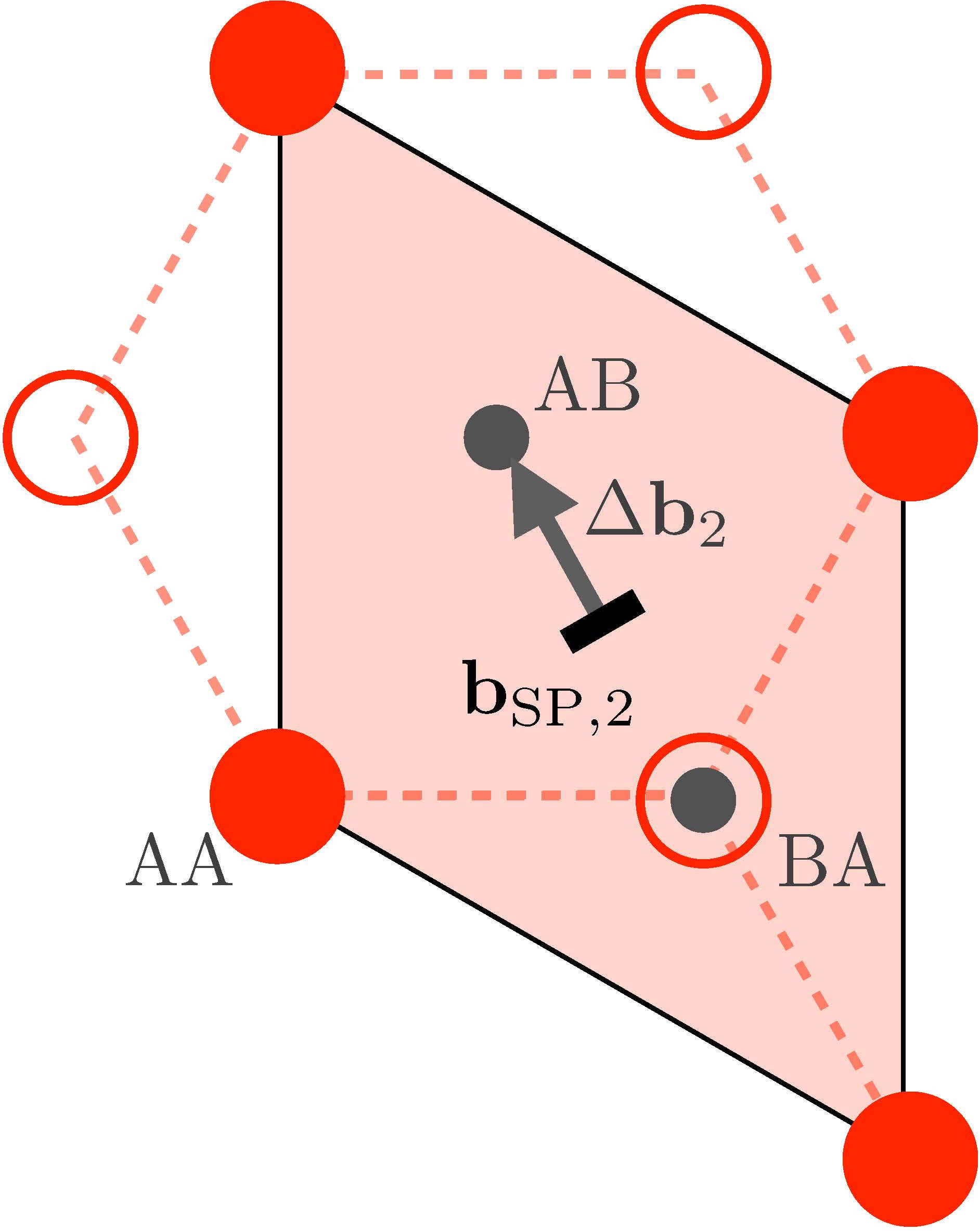}}
  \subcaptionbox{\label{fig:figure5c}}[.3\textwidth]
  {\includegraphics[height=2in]{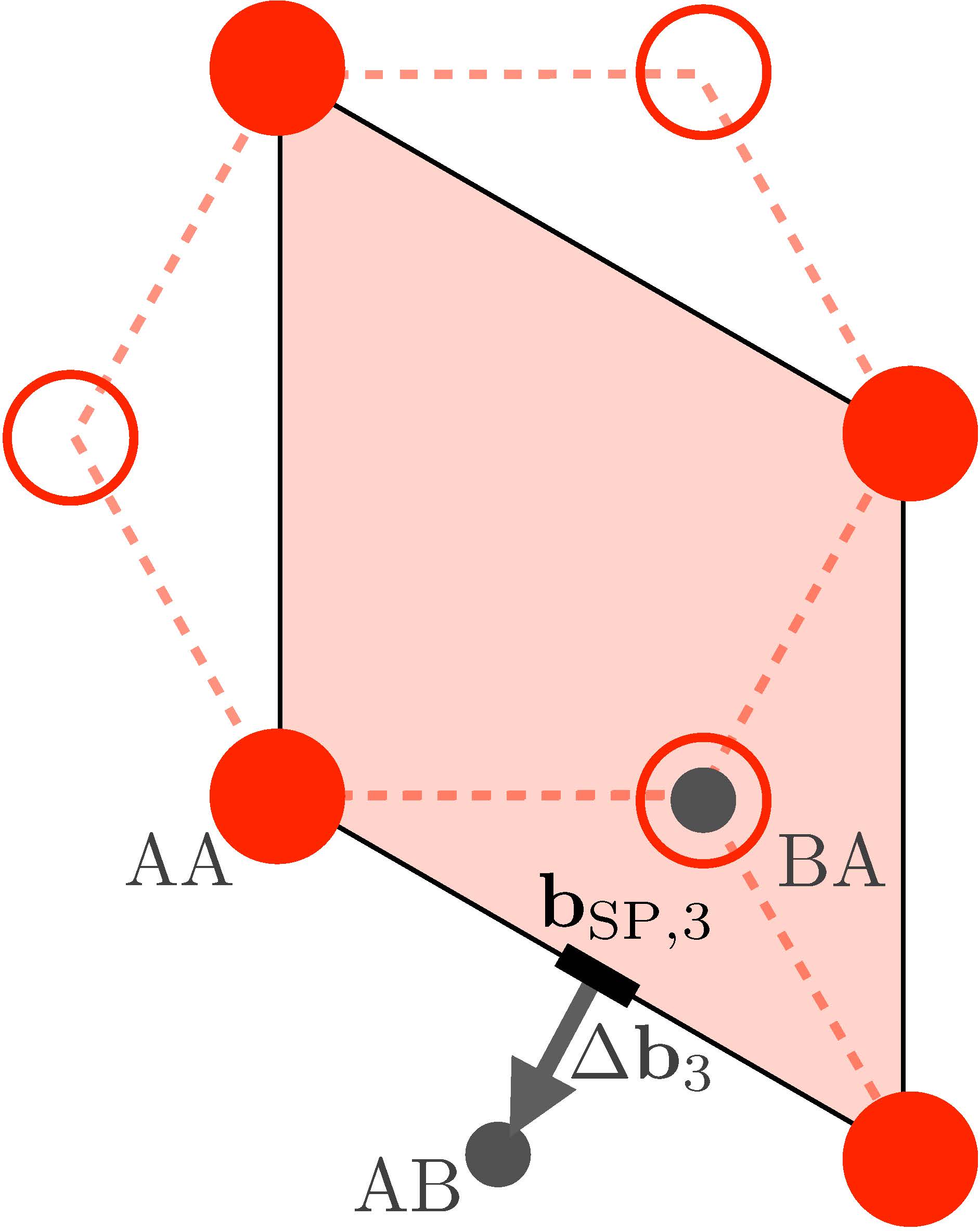}}
  \caption{Three possible orientations of the disregistry path across a domain wall for hexagonal bilayers. }
  \label{fig:figure5}
\end{figure}
\begin{figure}
    \centering
    \subcaptionbox{\centering Line cut along unit cell diagonal, with quartic and sine-Gordon fit shown.\label{fig:figure6a}}[.45\textwidth]
    {\includegraphics[width=.45\textwidth]{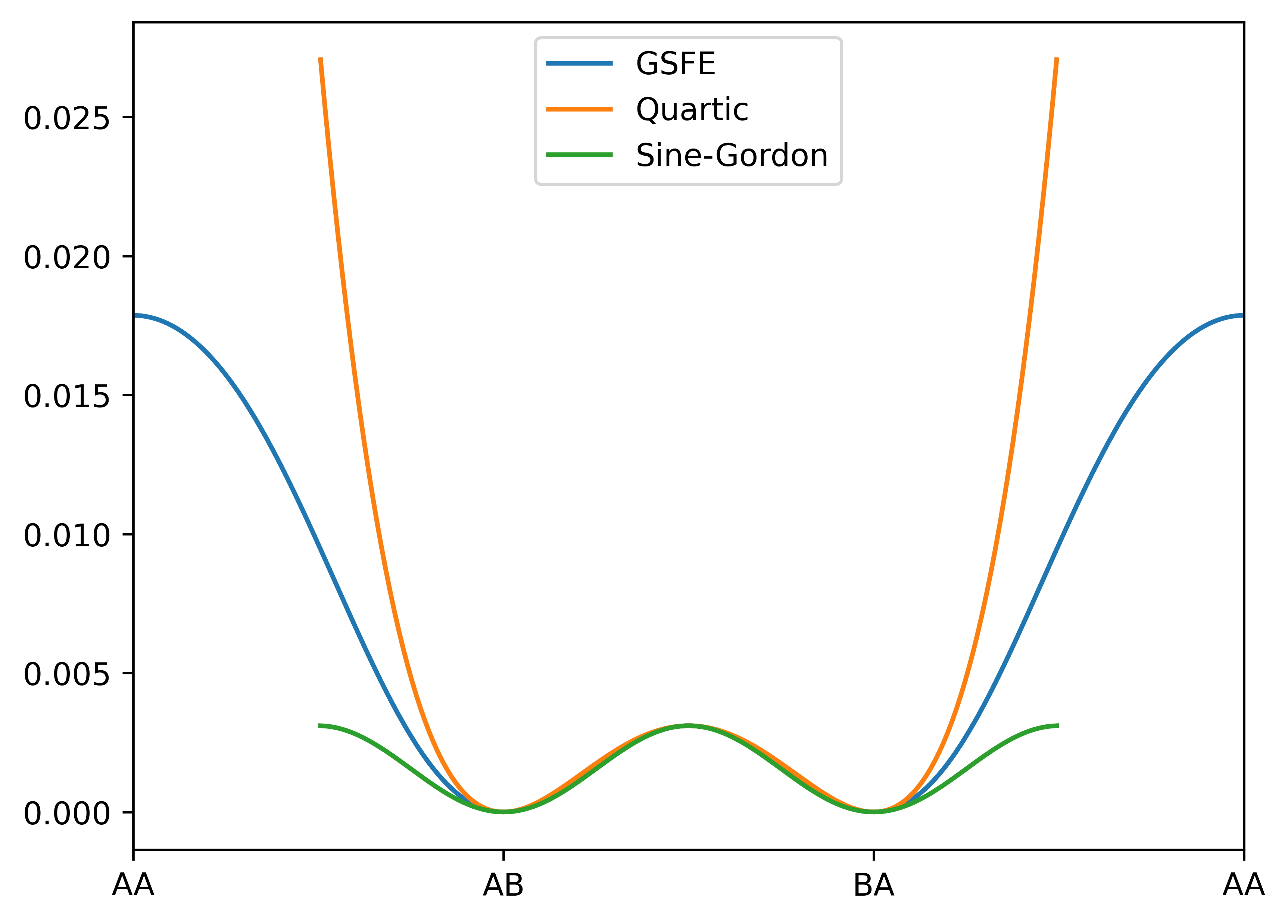}}
    \subcaptionbox{\centering Close-up between the AB and BA sites.\label{fig:figure6b}}[0.45\textwidth]
    {\includegraphics[width=.45\textwidth]{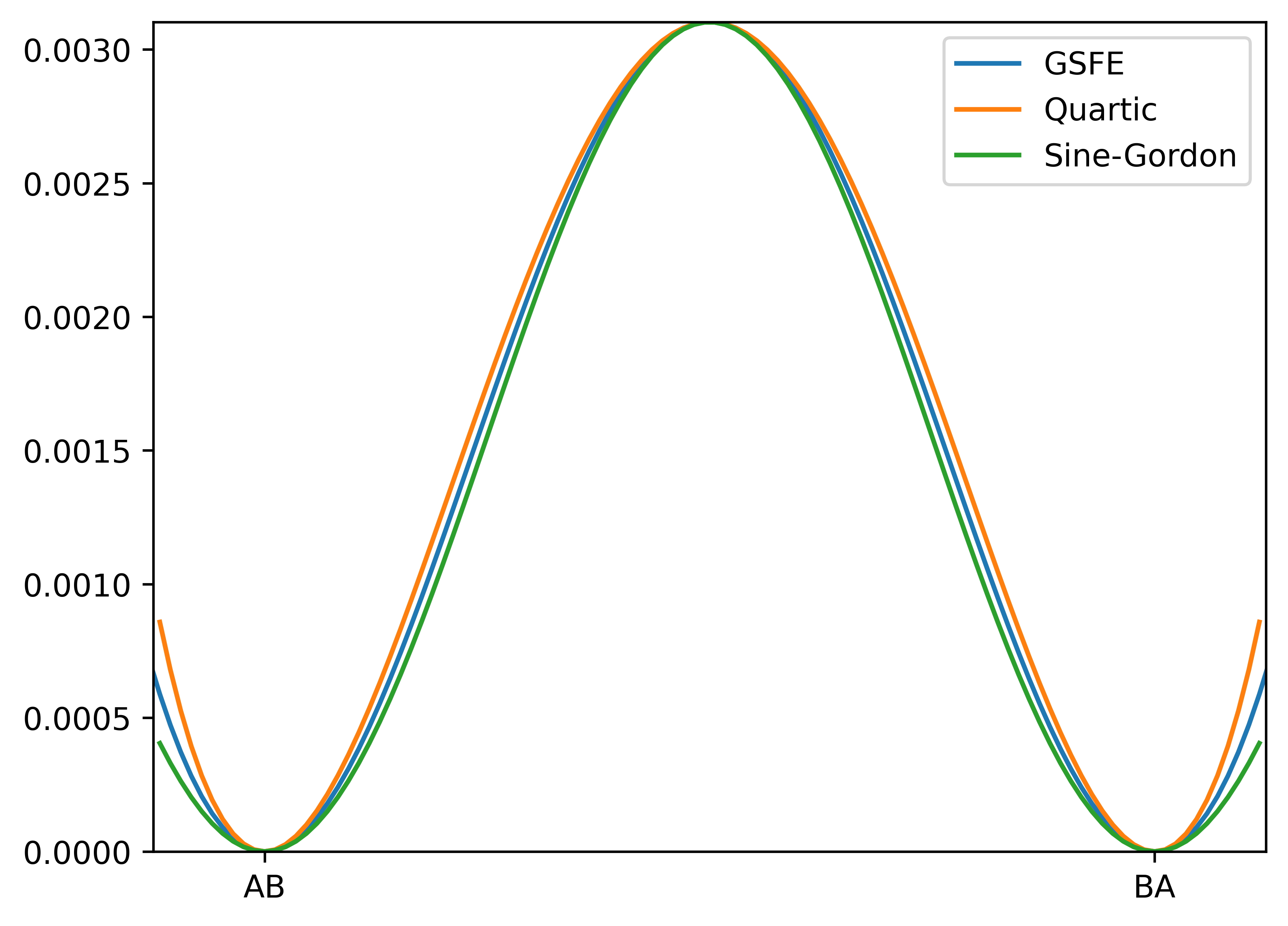}}
    \caption{GSFE potential profile along domain wall, comparisons to analytic potentials}\label{fig:figure6}
\end{figure}

The disregistry $\Delta \bb$ between left and right half-planes is chosen such that the unrotated GSFE profile  has minima at $\bb_\textrm{SP} \pm \Delta \bb$ and a saddle point at $\bb_\textrm{SP}$, leading to a double well profile on a segment $[\bb_\textrm{SP} - c \Delta \bb, \bb_\textrm{SP} + c  \Delta \bb]$ for some $c > 1$.
In the case of bilayer graphene with lattice given by~\eqref{def:GrapheneLattice}, such structures appear at the interface between Bernal stacked arrangements, with three non translationally equivalent choices corresponding to the $2\pi/3$ rotational symmetry of the lattice:
\[
  \begin{cases}
    \bb_{\textrm{SP},1} &= \dfrac{1}{2} a_0 \begin{pmatrix} 0 \\ 1 \end{pmatrix}, \\
    \Delta \bb_1 &= \dfrac{\sqrt{3}}{6}a_0 \begin{pmatrix} 1 \\ 0 \end{pmatrix},
  \end{cases} \qquad
  \begin{cases}
    \bb_{\textrm{SP},2} &= \dfrac{1}{2} a_0  \begin{pmatrix} \sqrt{3}/2 \\ 1/2 \end{pmatrix}, \\
    \Delta \bb_2 &= \dfrac{\sqrt{3}}{6}a_0 \begin{pmatrix} -1/2 \\ \sqrt{3}/2 \end{pmatrix},
  \end{cases} \qquad
  \begin{cases}
    \bb_{\textrm{SP},3} &= \dfrac{1}{2} a_0 \begin{pmatrix} \sqrt{3}/2 \\ -1/2 \end{pmatrix}, \\
    \Delta \bb_3 &= \dfrac{\sqrt{3}}{6}a_0 \begin{pmatrix}  -1/2 \\ -\sqrt{3}/2 \end{pmatrix},
  \end{cases}
\]
such that for $i = 1,2,3$, $\bb_\textrm{AB} = \bb_{\textrm{SP}, i} + \Delta \bb_i$ corresponds to AB stacking, while $\bb_\textrm{BA} = \bb_{\textrm{SP}, i} - \Delta \bb_i$ corresponds to BA stacking. The disregistry path associated with each is presented in Figure~\ref{fig:figure5}. For example, we note that going from the BA configuration (3) to the AB configuration (1) on Figure~\ref{fig:figure3} corresponds to the third choice here.

\subsection{Domain Wall Relaxation}
One expects the formation of a domain wall  at such an interface between stable domains, modeled by an interlayer displacement which we may assume depends only on the transverse coordinate $y$ and is aligned with the vector $R_\phi \Delta \bb$. By symmetry between the layers~\eqref{eq:interlayersym}, we obtain:
Since the disregistry is $\bb_{1 \to 2}(\bx)+ \bu_1(\bx)-\bu_2(\bx)=\bb_{1 \to 2}(\bx)+ 2\bu(\bx),$ we define the scalar interlayer displacement $u(y)$ by
\[
  \bu(\bx) := \bu_1(\bx) =  \frac{u(y)}2\, R_\phi \Delta \bb, \qquad \bu_2(\bx)= - \frac{u(y)}2\, R_\phi \Delta \bb.
\]

Plugging this ansatz into the minimization problem~\eqref{eq:genPN_sym}, we replace the integration of the energy over the moiré cell $\Gamma_\mathcal{M}$ by an average energy per unit length along the domain wall. This yields the simplified energy minimization problem for domain wall relaxation:
\begin{equation}\label{eq:DWrelax}
  \mathcal{E}(u) = \int_{-\infty}^\infty \frac{1}{2} \left (  \frac{\lambda+\mu}{2}  (R_\phi \Delta \bb \cdot \be_y)^2 + \frac{\mu}{2} \Vert \Delta \bb \Vert^2 \right ) u_y^2(y) + \Phi \left [u(y) \right ],
\end{equation}
with the boundary conditions $u(y) \to \pm 1$ for $y \to \pm \infty$, where we introduced the effective one-dimensional, angle-independent GSFE potential:
\[
    \Phi[u] := \Phi_1[R_\phi ( \bb_{\textrm{SP}} + u \Delta \bb )] = \Phi_2[\ml{-} R_\phi (\bb_{\textrm{SP}} + u \Delta \bb )].
\]
We recall that $\lambda+\mu$ is the bulk modulus and $\mu$ is the shear modulus.
The potential $\Phi$ has a double-well even profile over the the interval $[-c,c]$, with minimal value $\Phi_\textrm{min}$ attained at $u = \pm 1$, and $k_\textrm{min} = \Phi''(-1) = \Phi''(1) > 0$.

We further non-dimensionalize the problem as follows: let us define the order parameter $\psi$ by rescaling position with respect to the characteristic width $l_\phi$:
\begin{equation}\label{eq:DWcharacteristics}
  \psi (t) = u(l_\phi\, t),
  \qquad
  \text{where}
  \quad
  l_\phi = \sqrt{ \frac{\left ( \lambda + \mu \right ) \cos(\theta_0 + \phi)^2 + \mu}{2 k_\textrm{min}}  } \; \Vert \Delta \bb \Vert,
\end{equation}
with $\theta_0 + \phi$ is the angle between the interlayer translation direction $R_\phi \Delta \bb$ and the domain wall normal ($\theta_0=0$ for Fig \ref{fig:figure5}a,
$\theta_0=2\pi/3$ for Fig \ref{fig:figure5}b, and $\theta_0=4/3\pi$ for Fig \ref{fig:figure5}c), and the normalized Ginzburg-Landau-type potential:
\[
    U(\psi) = \frac{1}{k_\textrm{min}} \left ( \Phi\left [ \psi \right ]  - \Phi_\textrm{min} \right )\geq 0.
\]
From the Euler-Lagrange equation associated with the minimization problem~\eqref{eq:DWrelax}, we obtain the  non-dimensionalized, one-dimensional relaxation boundary value problem with the angle-independent order parameter profile $\psi$ interpolating between the two minima $\pm 1$ of the double well potential $U(\psi)$:
\begin{equation}\label{eq:DW}
   \left \{ \begin{aligned}
    & \psi'' = U'(\psi), \\
    &\psi(t) \to \pm 1 \qquad \text{as } t \to \pm \infty.
  \end{aligned} \right.
\end{equation}

The minimization problem with double well profile~\eqref{eq:DWrelax} as well as the associated Euler-Lagrange equation~\eqref{eq:DW} are classical examples exhibiting topological soliton or kink solutions, with typical examples being the Allen-Cahn (or quartic) or sine-Gordon equations for which an analytic formula for the solution is available. Under mild regularity conditions on the GSFE functional $\Phi$, we have the following classical result (see e.g.~\cite[Prop. 2.1]{jendrej2019dynamics}:
\begin{proposition}\label{prop:dw}
    Assume $U(\psi)$ is even, twice continuously differentiable, with global minimum $U(\pm 1) = 0$, $U > 0$ for $-1 < \psi < 1$ and $U''(\pm 1) = 1 $. Then, there exists a unique profile $\psi(t)$ solution of~\eqref{eq:DW} such that $\psi(0) = 0$. In addition, we have the asymptotic estimates with $\kappa, C>0$:
    \[
        \vert \psi(t) + 1 - \kappa e^{t} \vert \leq C e^{2t}, \qquad \vert \psi(t) - 1 + \kappa e^{-t} \vert \leq C e^{-2t} \qquad \forall t \in \mathbb{R}.
    \]
\end{proposition}
\noindent Hence, the relaxation model predicts the formation of an exponentially localized domain wall with the profile $u(y) = \psi(y/l_\phi)$, such that the characteristic width $l_\phi$ is smallest for a shear
boundary (interlayer translation $R_\phi \Delta \bb$ parallel to the domain wall) with $l_\perp = \sqrt{\frac{\mu}{2 k_\textrm{min}}}  \Vert \Delta \bb \Vert$, and largest for a tensile boundary (interlayer translation orthogonal to the domain wall) with $l_\parallel = \sqrt{\frac{\lambda + 2\mu}{2 k_\textrm{min}}} \Vert \Delta \bb \Vert$. \pc{This prediction matches experimental measurements of domain wall thicknesses} \ml{reported in \cite[Figure 3H]{alden2013strain}}.\footnote{Note that in~\cite{alden2013strain} the GSFE profile across the domain wall is approximated by the sine-Gordon potential $\Phi(\Delta u) = \frac{V_\textrm{SP}}{2} (1 - \cos(2\pi \Delta u))$ for $0 \leq \Delta u \leq 1$ allowing for an explicit solution, and a different normalization was used with $k = \Phi[0] - \Phi_\textrm{min}$ the saddle point energy, resulting in a slightly different expression for the thickness of the domain wall.
}

\section{Four types of moiré lattices}\label{sec:4types}
When stacking two layers of the same material on top of each other, a large-scale moiré pattern
\begin{wrapfigure}{r}{.5\textwidth}
\centering \vspace{-.1in}
\includegraphics[width=.42\textwidth]{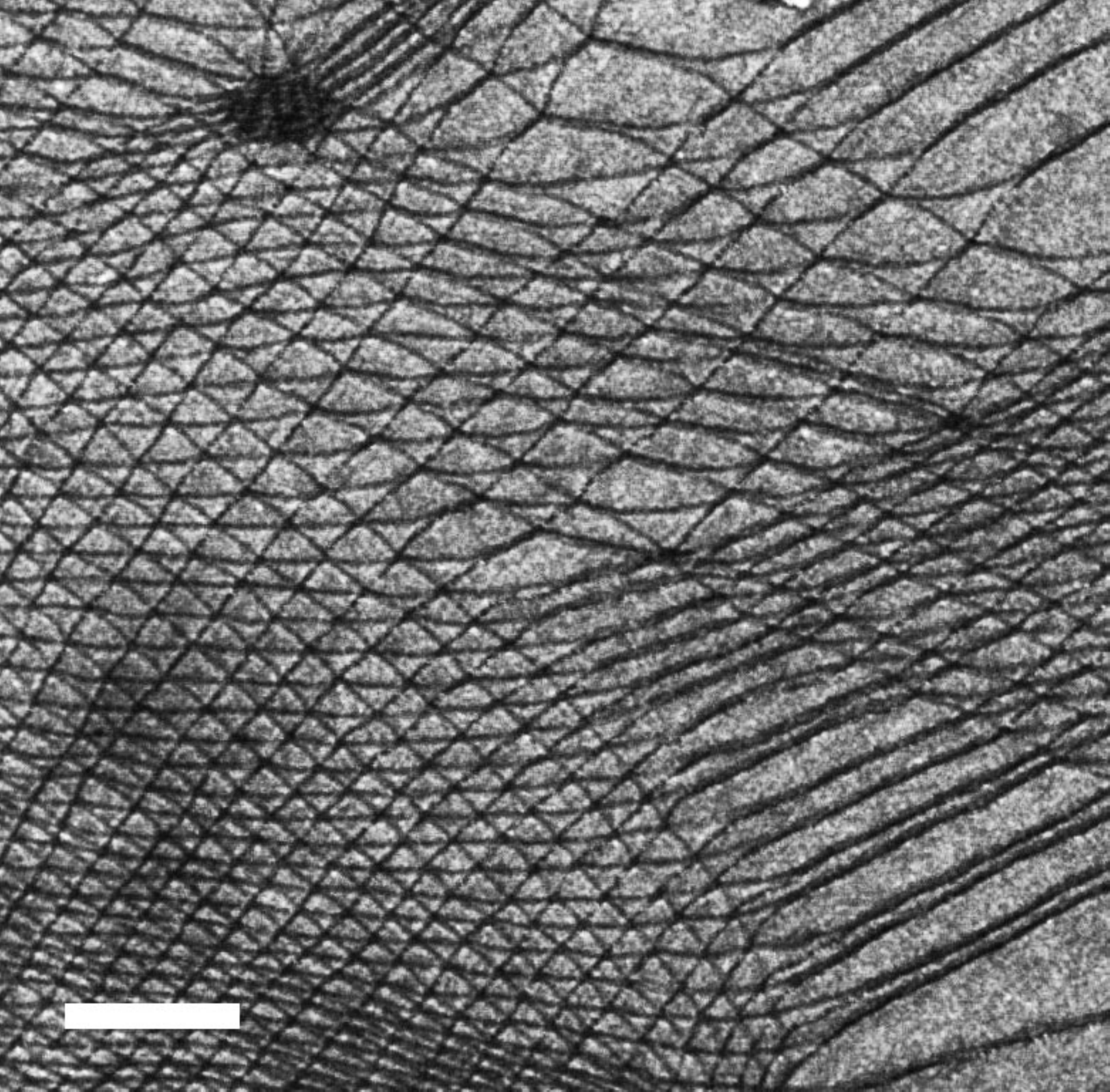}
\caption{Experimental Dark Field Image of a MoSe$_2$/WSe$_2$ heterostructure exhibiting a variety of moiré patterns due to a non-uniform strain field.}
\label{fig:figure7}
\vspace{-.4in}
\end{wrapfigure}
may emerge due to a small rotation or deformation of one or both layers.
In this section, we present the ``pure'' cases of such global patterns: first, the well-studied \textit{twisted} bilayer case; next, isotropic strain, pure shear, and simple shear.
We observe from experimental images, Figure~\ref{fig:figure1} and Figure~\ref{fig:figure7} that each of these cases can exist in a non-uniformly stacked bilayer structure and the results in this section on the orientation of the moir\'e patterns can guide the interpretation of the local stacking.

\begin{figure}[ht]
  \centering
  \subcaptionbox{Twist ($\theta = 8^\circ$)\label{fig:figure8a}}[.45\textwidth]
  {\includegraphics[width=.45\textwidth]{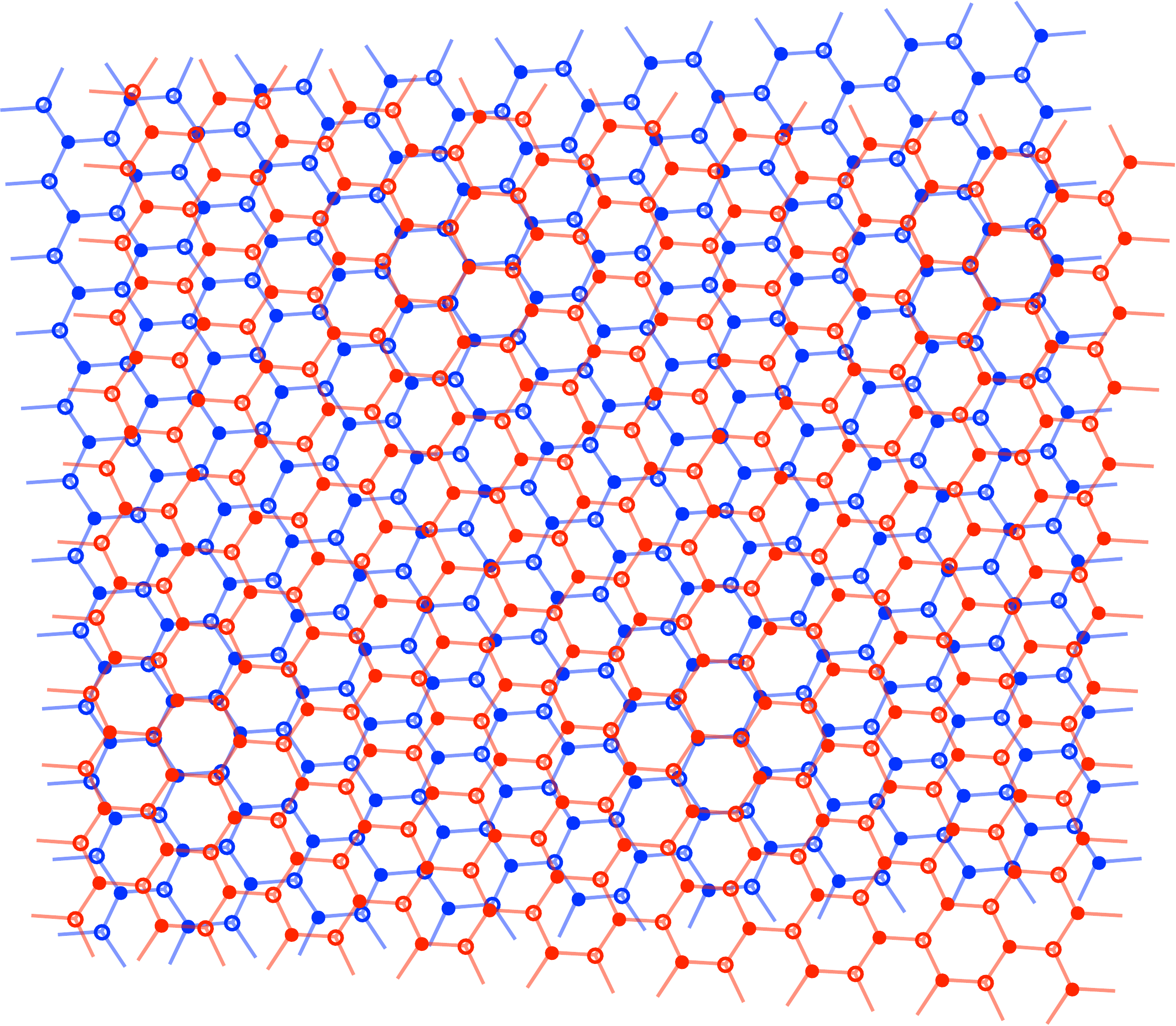}}
  \subcaptionbox{Isotropic strain ($\varepsilon = 10\%$)\label{fig:figure8b}}[.45\textwidth]
  {\includegraphics[width=.45\textwidth]{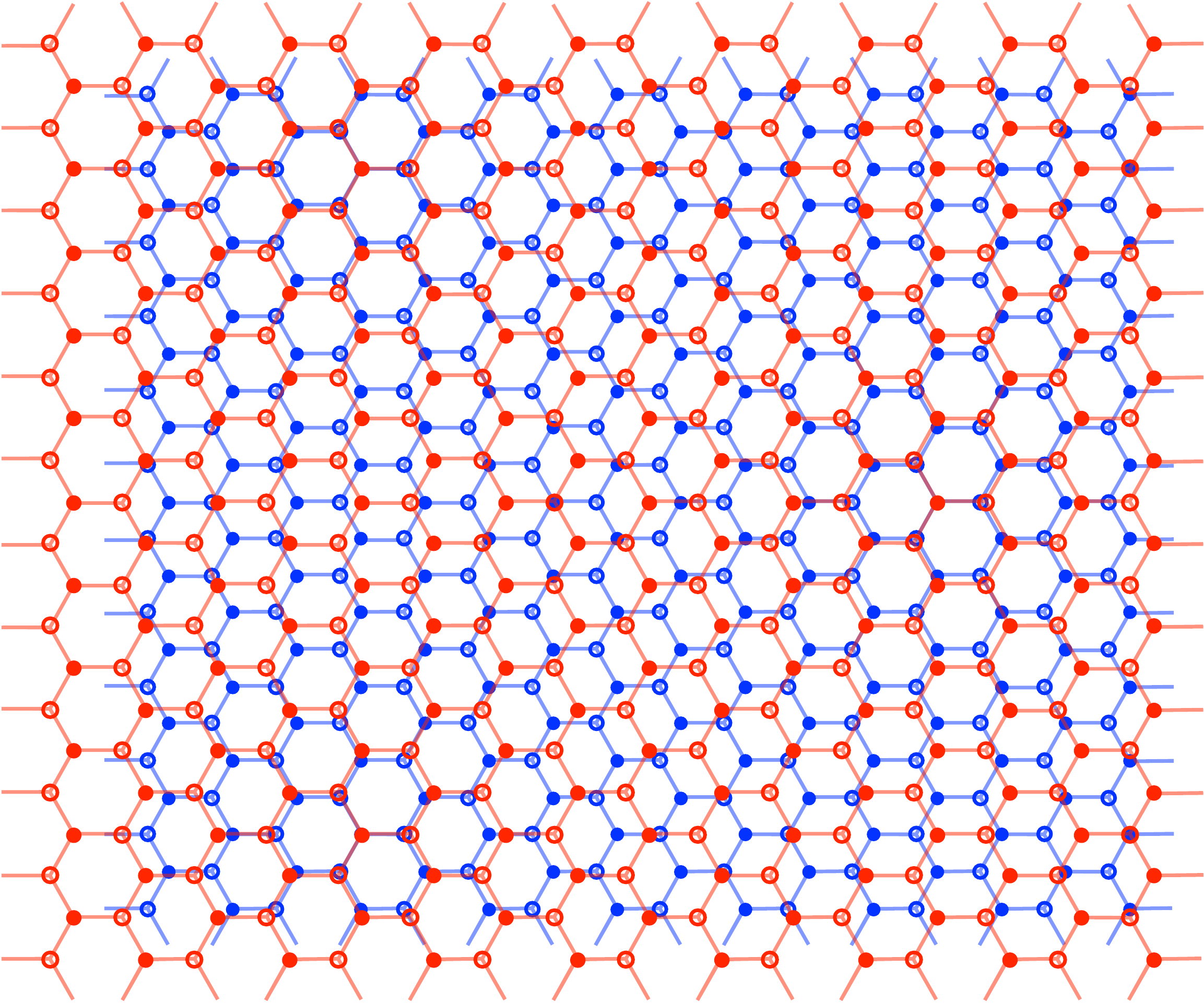}}
  \subcaptionbox{Diagonal pure shear ($\varepsilon = 10\%$)\label{fig:figure8c}}[.45\textwidth]
  {\includegraphics[width=.45\textwidth]{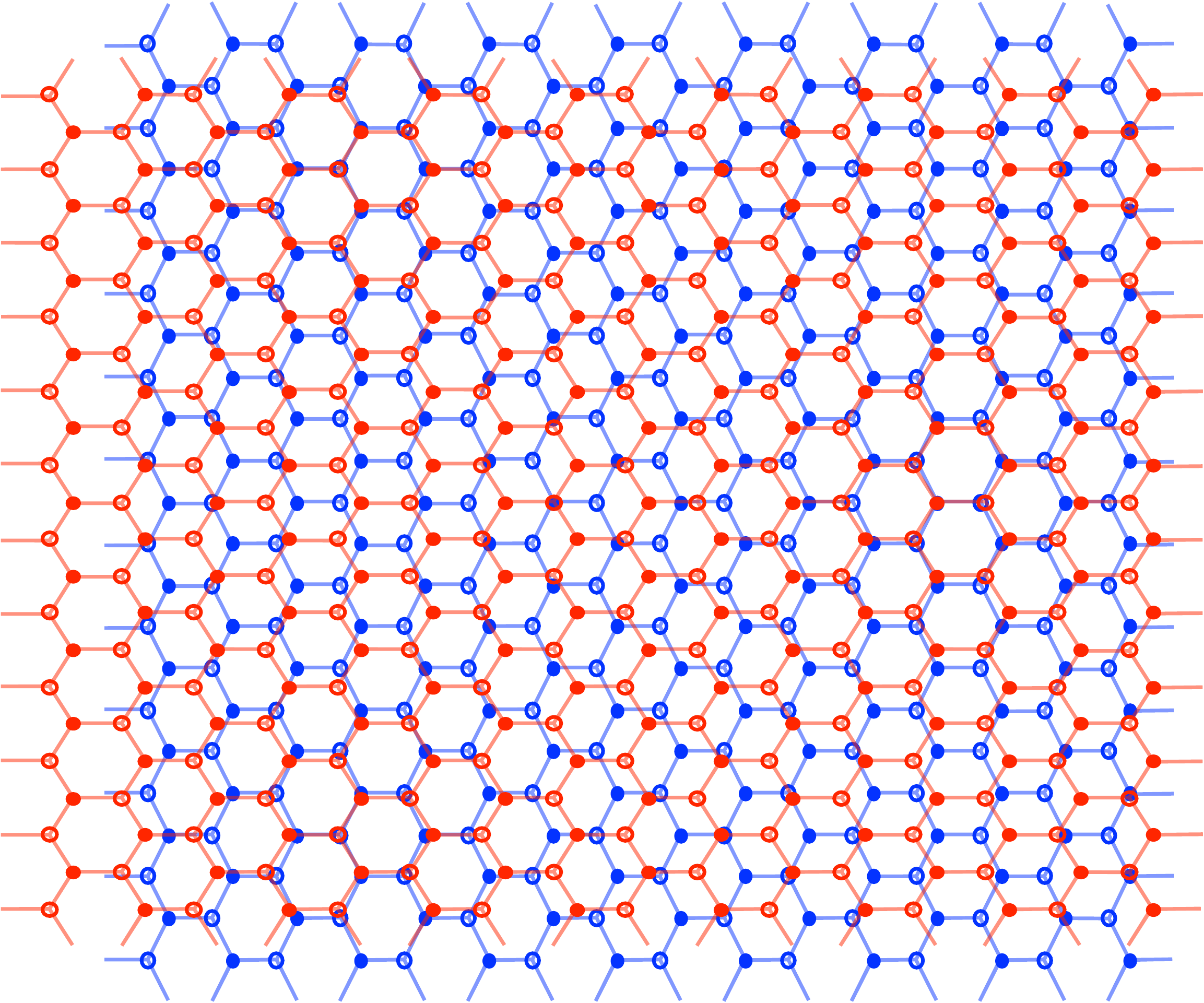}}
  \subcaptionbox{Horizontal simple shear ($\varepsilon = 10\%$)\label{fig:figure8d}}[.45\textwidth]
  {\includegraphics[width=.45\textwidth]{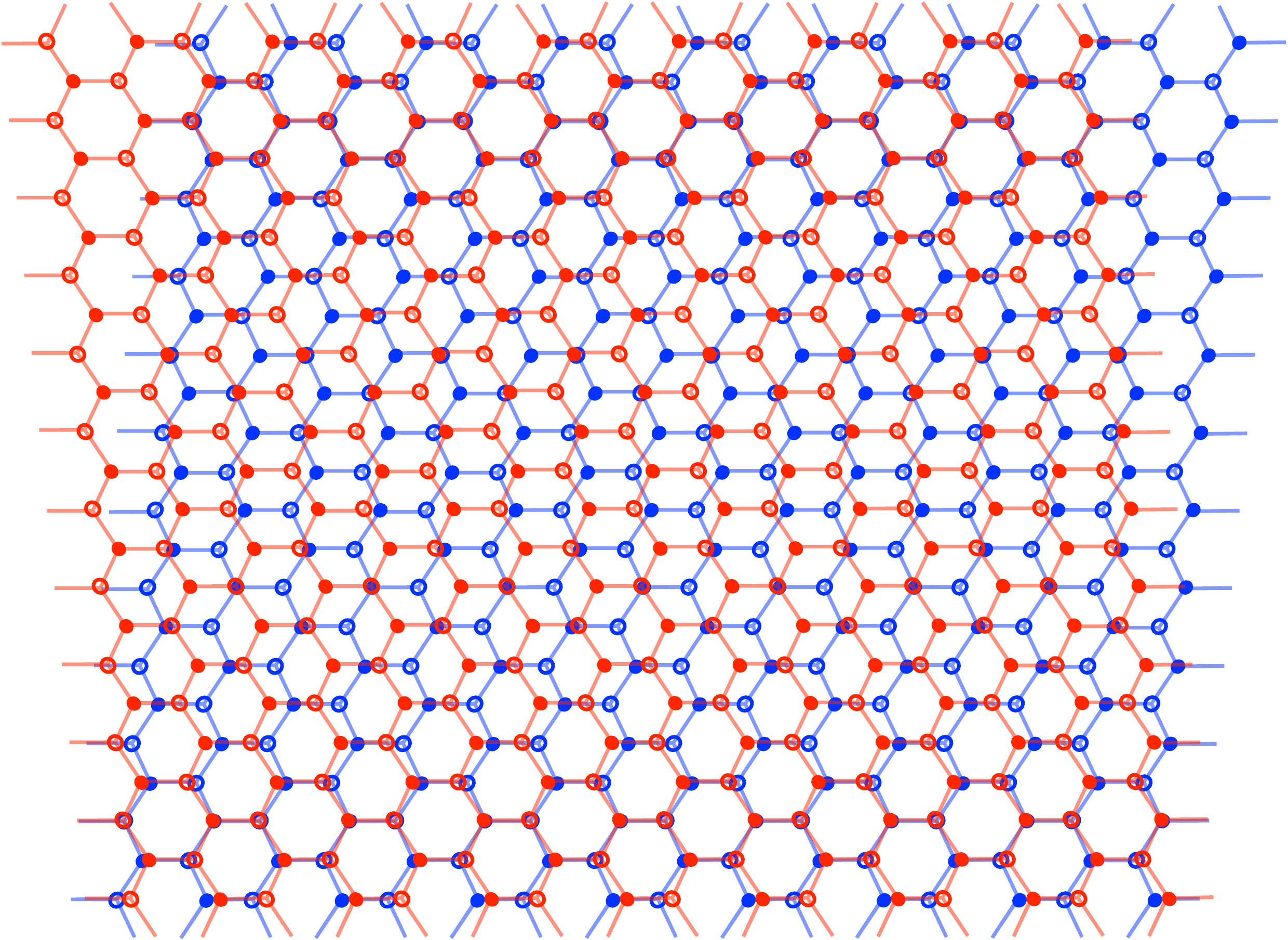}}
  \caption{Moiré patterns obtained by applying a small uniform strain to two hexagonal layers.}
  \label{fig:figure8}
\end{figure}

\subsection{Twist}
The layers are respectively twisted by an angle $\theta$:
\[
    A_1 = R_{-\theta/2} A, \qquad A_2 = R_{+\theta/2} A.
\]
This case has been extensively studied both experimentally~\cite{KimRelax18} as well as theoretically~\cite{dai2016twisted,relaxphysics18,cazeaux2020energy}.
The moiré pattern in this case is a triangular lattice with fundamental matrix~\eqref{def:moirelattice}
\[
    A_\mathcal{M} = \frac{1}{2 \sin(\theta/2)} J\, A \qquad \text{where} \quad J = 
    R_{-\pi/2}=\begin{bmatrix} 0 & 1 \\ -1 & 0 \end{bmatrix},
\]
i.e., it is rotated by $-\pi/2$ from the reference lattice, and scaled inversely proportionally to the twist angle as seen on Figure~\ref{fig:figure8a}.
We can similarly calculate that
\begin{align*}
    {\mathbf{b}}_{1 \to 2}:
    \begin{cases} \Gamma_\mathcal{M} \to \Gamma_2, \\ \mathbf{x} \mapsto 2\sin (\theta/2) J R_{\theta/2} \bx, \end{cases}
    \qquad {\mathbf{b}}_{2 \to 1}:
    \begin{cases} \Gamma_\mathcal{M} \to \Gamma_1, \\ \mathbf{x} \mapsto -2\sin (\theta/2) J R_{-\theta/2} \bx. \end{cases}
\end{align*}

By symmetry of the unrelaxed twisted bilayer structure, there exists a single GSFE functional
$\Phi_0(\bx) := \Phi_{1}({\mathbf{b}}_{1 \to 2}(\bx)) = \Phi_{2}({\mathbf{b}}_{2 \to 1}(\bx))$
where
$\Phi_0:\Gamma_\mathcal{M}\to \R.$
We can then derive the local interlayer energies with relative displacements $\bv(\bx)$ to be
\begin{equation*}
    \begin{split}
        \Phi_{1}({\mathbf{b}}_{1 \to 2}(\bx) + \bv(\bx)) &+ \Phi_{2}({\mathbf{b}}_{2 \to 1}(\bx) - \bv(\bx)) \\&=
    \Phi_0(\bx  - (2\sin(\theta/2))^{-1}J R_{-\theta/2} \bv) + \Phi_0(\bx  - (2\sin(\theta/2))^{-1}J R_{\theta/2} \bv)
    \end{split}
\end{equation*}
and the total interlayer energy to be
\begin{equation}\label{eq:misfitEnergytwist}
    \mathcal{E}_\mathrm{inter}[\bv] = \frac{1}{2} \int_{\Gamma_\mathcal{M}} \Phi_0(\bx  - (2\sin(\theta/2))^{-1}J R_{-\theta/2} \bv) + \Phi_0(\bx  - (2\sin(\theta/2))^{-1}J R_{\theta/2} \bv)\, d \bx.
\end{equation}

\begin{figure}[ht!]
    \centering
    \subcaptionbox{Unrelaxed configuration\label{fig:figure9a}}[.49\textwidth]{\includegraphics[width=.45\textwidth]{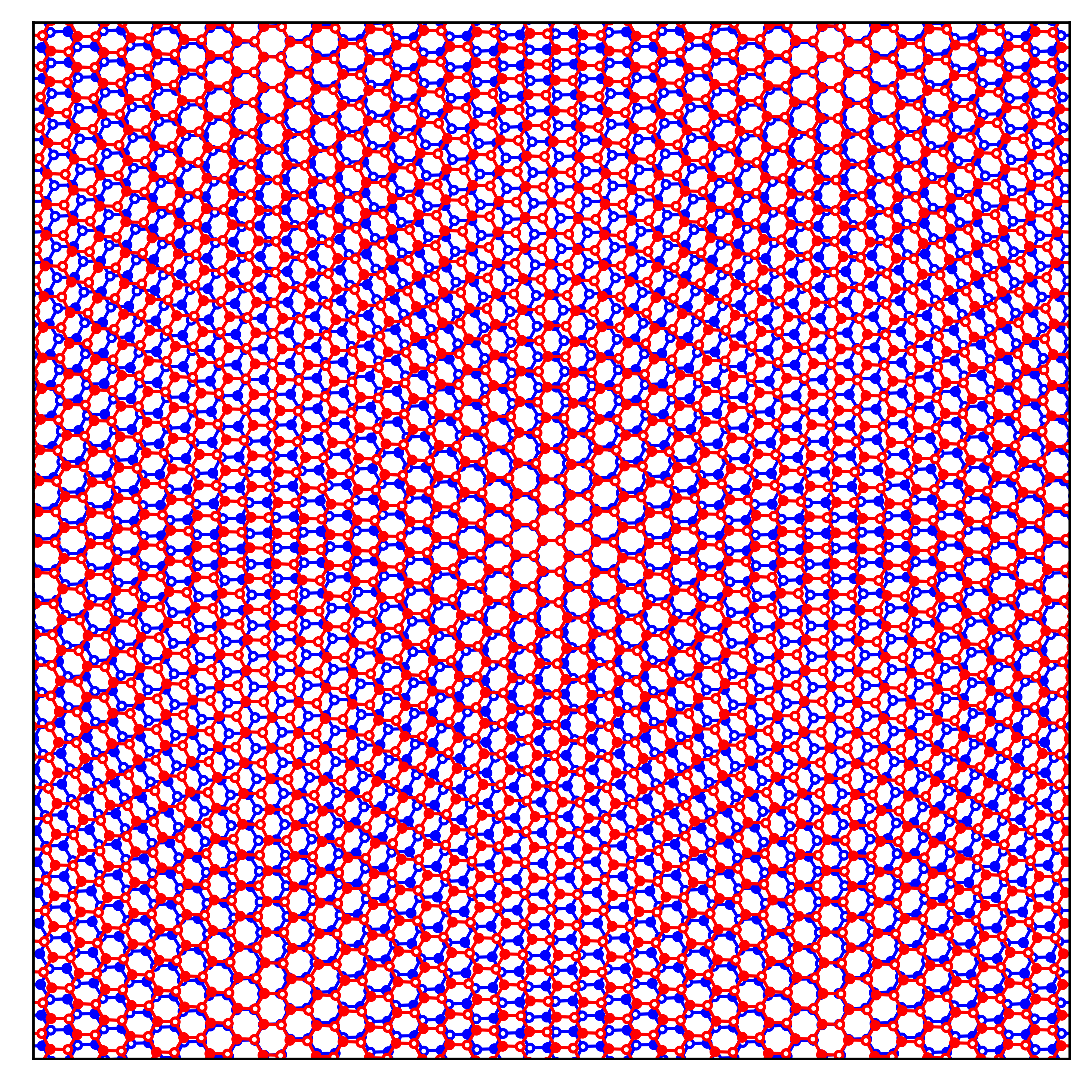}} \\
    \subcaptionbox{Relaxed configuration\label{fig:figure9b}}[.49\textwidth]{\includegraphics[width=.45\textwidth]{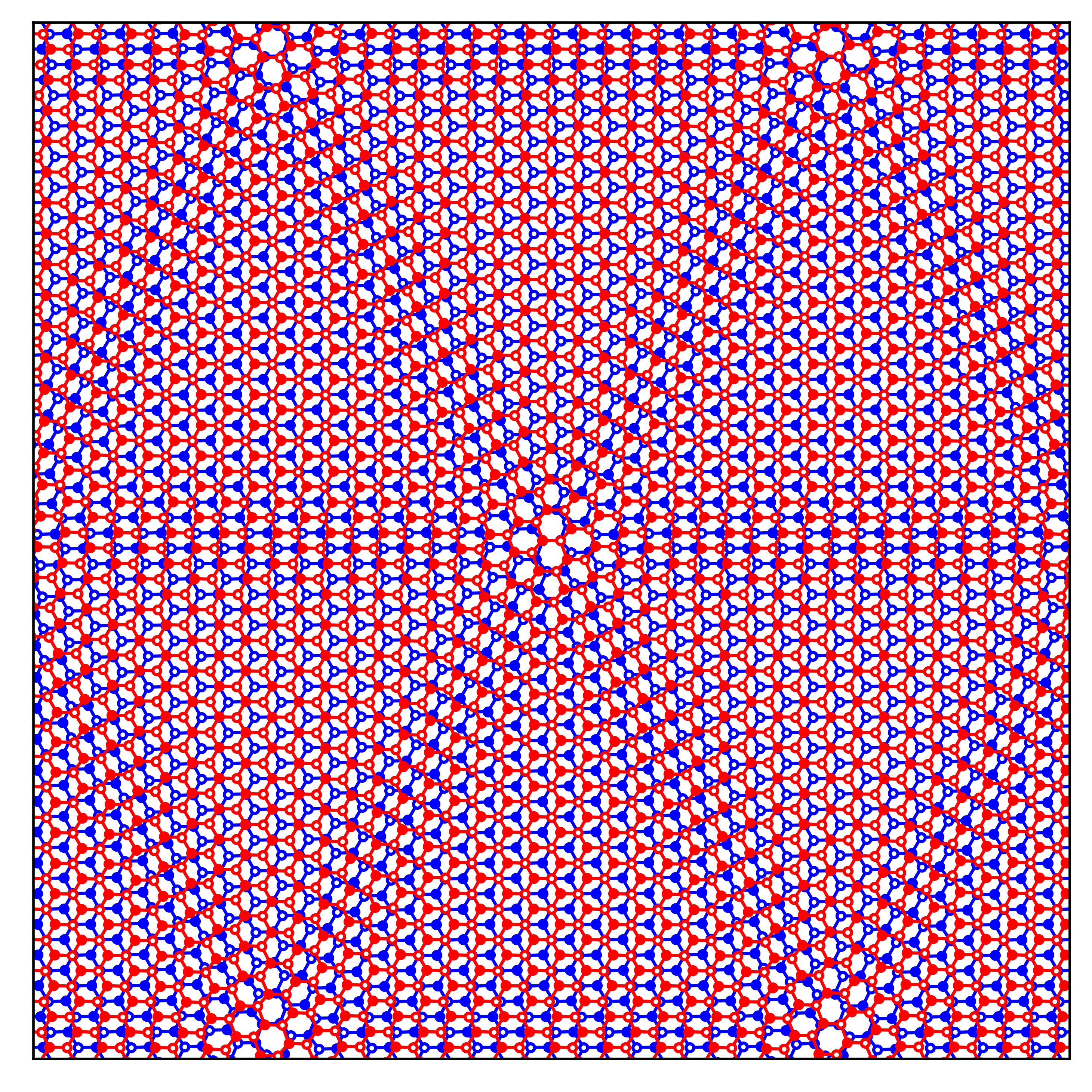}}
    \subcaptionbox{Schematic description of domain formation in twisted bilayer structures\label{fig:figure9c}}[.49\textwidth]
    {\includegraphics[width=.45\textwidth]{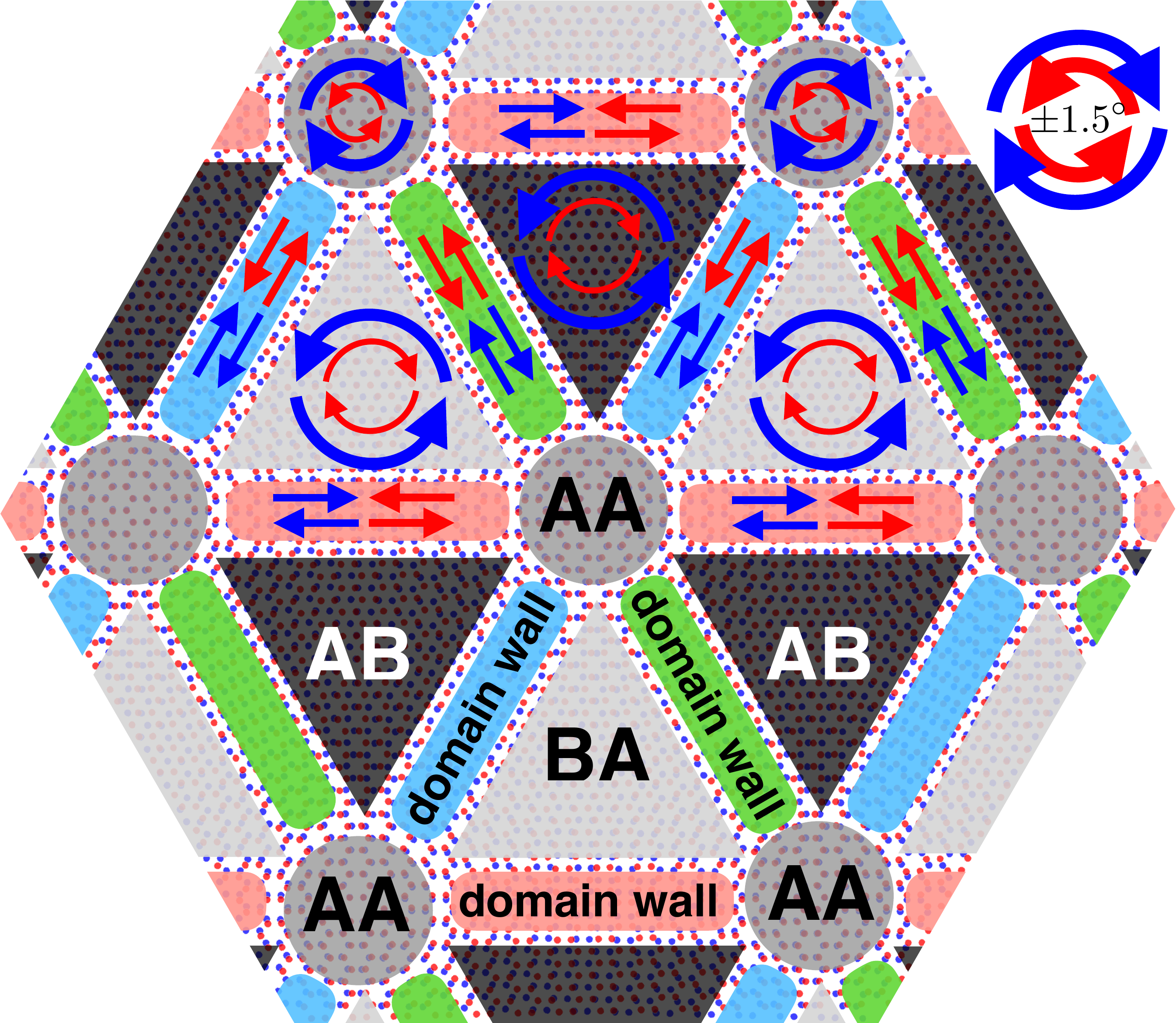}\vspace{.2in}}
    \caption{Twist moiré ($\theta=3.1^\circ$) with misfit energy amplified by a factor of 100.}
    \label{fig:figure9}
\end{figure}

Following~\cite{cazeaux2020energy}, we can perform an asymptotic study by introducing a new angle-independent reference cell $\Gamma_0$ such that
\begin{equation}\label{eq:}
    \Gamma_\mathcal{M}(\theta) = \frac{1}{2\sin(\theta/2)} \Gamma_0, \quad \text{where} \
    \begin{cases} \mathcal{R}_0 := A_0 \mathbb{Z}^2, \\ \Gamma_0 := \mathbb{R}^2 / \mathcal{R}_0, \end{cases}
    \quad \text{with } A_0 := J A.
\end{equation}
We can then obtain by the scaling $\widehat{\bx} = 2\sin(\theta/2) \bx $ and $\Psi_0(\widehat{\bx})=\Phi_0((2\sin(\theta/2))^{-1}\widehat{\bx})=\Phi_0(\bx)$
the relaxation problem~\eqref{eq:genPN_sym} recast as a problem on the fixed reference cell $\Gamma_0$:
\begin{equation}\label{eq:twistPN}
    \text{Minimize} \quad \mathcal{E}_0^\theta[\bu] := 2 \mathcal{E}^0_\mathrm{intra}[\bu] + \mathcal{E}^\theta_\mathrm{inter}[2\bu],
\end{equation}
where 
\begin{align*}
    \mathcal{E}^0_\mathrm{intra}[\bu] &:= \int_{\Gamma_0} \frac{\lambda}{2} (\mathrm{div}\ \bu)^2 + \mu \varepsilon(\bu) : \varepsilon(\bu), \\
    \mathcal{E}^\theta_\textrm{inter}[\bv] &:= \frac{1}{4\sin^2(\theta/2)} \int_{\Gamma_0} \frac 12 \left ( \Psi_0(\widehat{\bx}  - J R_{\theta/2} \bv) + \Psi_0(\widehat{\bx} - J R_{-\theta/2} \bv) \right ) d\widehat{\bx}.
\end{align*}
This explicit rescaling allows us to state the following result, improving on previous results in \cite{cazeaux2020energy} by a simple variational argument:
\begin{theorem}\label{thm:est}
    Assume that the functional $\Psi_0$ is twice continuously differentiable. Then for any angle $\theta$ which is not an integer multiple of $\pi$, there exists periodic minimizers $\bu^*$ to the relaxation problem~\eqref{eq:twistPN} in the Sobolev space $W^{1,2}_0(\Gamma_0) := \{ \bu \in W^{1,2}_0(\Gamma_0) \ \vert \ \int_{\Gamma_0} \bu(\widehat{\bx}) d \widehat{\bx}  = {\boldsymbol 0} \}$, which satisfy
    \[
        \Vert \bu^* \Vert_{1,2} \leq \frac{1}{2 \sin(\theta/2)} \left ( \frac{1}{\mu C_A^1} \int_{\Gamma_0} [\Psi_0(\bx) - \Psi_{\mathrm{min}} ] d\widehat{\bx}  \right )^{1/2},
    \]
    where $C^1_A$ is a fully identifiable constant depending on the lattice basis $A$ only.
\end{theorem}
This new estimate shows in particular that the $L^2$-norm of the gradient on the moiré unit cell scales at most \textit{inversely linearly} with the twist angle, i.e., linearly with the width of the moiré domains, a result consistent with the formation of one-dimensional domain walls with a fixed width around triangular domains at very small twist angles, as observed in simulations (see~Figure \ref{fig:figure9}) and experiments~\cite{KimRelax18}.
\begin{proof}
    As noted in Remark 4.5 and Theorem 4.10 in~\cite{cazeaux2020energy}, there exists global minimizers $\bu^*$ in $W^{1,2}_0(\Gamma_0)$ to problem~\eqref{eq:twistPN} which solve the associated Euler-Lagrange equation and satisfy the following bound:
    \[
        \Vert \bu^* \Vert_{1,2} \leq \frac{1}{\mu C_A^1} \frac{\Vert \nabla \Psi_0 \Vert_\infty }{4 \sin^2(\theta/2)},
    \]
    where $C_A^1$ is a constant such that the following Korn-type inequality is satisfied for all functions $u \in W^{1,2}_0(\Gamma_0)$:
    \[
        \int_{\Gamma_0} \lambda (\div \bu)^2 + 2 \mu \varepsilon(\bu) : \varepsilon(\bu) d\widehat{\bx} \geq \mu C_A^1 \Vert \bu \Vert_{1,2}^2.
    \]
    On the other hand, we note that the energy bound
    \[
        \mathcal{E}_0^\theta [\bu^*] \leq \mathcal{E}_0^\theta [\boldsymbol{0}] = \frac{1}{4\sin^2\theta/2} \int_{\Gamma_0} \Psi_0(\bx) d\widehat{\bx}
    \]
    directly implies
 \begin{equation*}
  \begin{split}
        \mu C_A^1 \Vert \bu^* \Vert_{1,2}^2 &\leq \int_{\Gamma_0} \lambda (\div \bu^* )^2 + 2 \mu  \varepsilon(\bu^*) : \varepsilon(\bu^*) d\widehat{\bx}
        = \mathcal{E}_0^\theta [\bu^*]-\mathcal{E}^\theta_\textrm{inter}[2\bu^*]\\
        &\leq  \frac{1}{4\sin^2\theta/2} \int_{\Gamma_0} [ \Psi_0(\widehat{\bx}) - \Psi_\mathrm{min} ] d\widehat{\bx}.
\end{split}
\end{equation*}

\end{proof}

\noindent To visualize the nature of the relaxed pattern in the case of twisted bilayers with a hexagonal lattice (such as graphene or other TMDs), we present unrelaxed (Fig.~\ref{fig:figure9a}) and relaxed (Fig.~\ref{fig:figure9b}) atomic moiré patterns, computed using the methods of Section~\ref{sec:numerical} to calculate displacements $\bu_1, \bu_2$ and displacing the atoms of layer $i = 1,2$ as $\bR \to \bR + \bu_{i}(\bR)$, using an exagerated misfit energy for illustration purposes.
\begin{figure}[hp!]
    \begin{minipage}{.37\textwidth}
        \subcaptionbox{
        Relaxed GSFE over twist angles $0.8^\circ$, $0.4^\circ$, $0.2^\circ$, and $0.1^\circ$ (from the top down) scaled and recentered to show the moir\'e scale (left). \dc{The $0.4^\circ$ plot also shows the normal to the domain wall (b) in orange.}
        }[\textwidth]
        {\includegraphics[width=\textwidth]{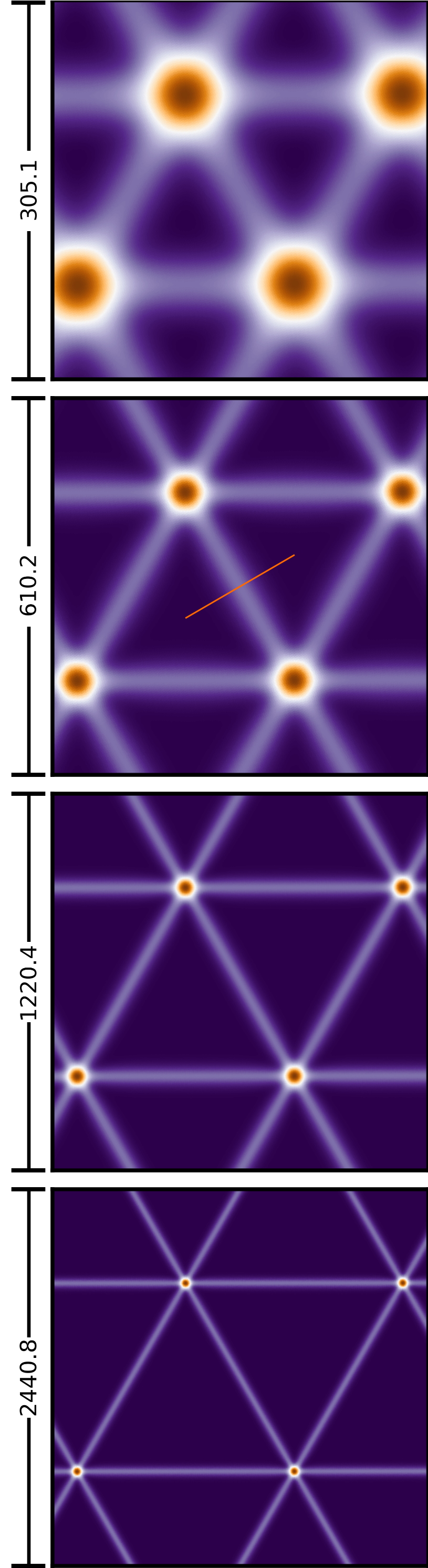}}
    \end{minipage}
    \hspace{.1\textwidth}
    \begin{minipage}{.5\textwidth}
        \subcaptionbox{
        Order parameter $\psi$~\eqref{eq:DWcharacteristics} plotted across an AB-BA domain wall for various angles. At~$0.8^\circ$ the plot is cut off beyond the AB and BA points.
            \label{fig:figure10b}
        }[\textwidth]
        {
        \includegraphics[width=\textwidth]{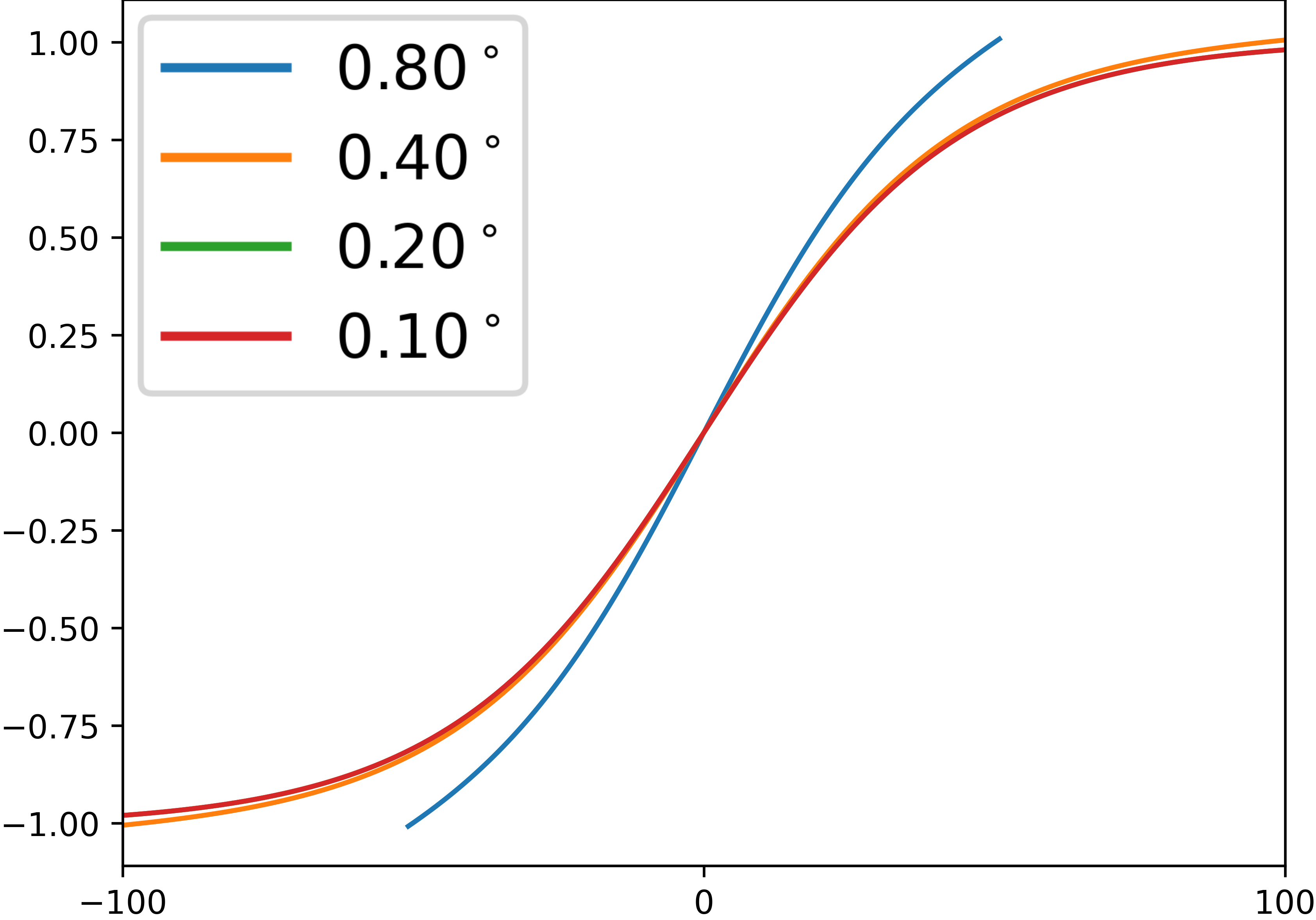}
        }
        \smallskip

        \subcaptionbox{
        Relaxed GSFE with angle of $0.2^\circ$ (top) and $0.1^\circ$ (bottom) plotted across the same region and a consistent scale.
        }[\textwidth]
        {
        \includegraphics[width=.95\textwidth]{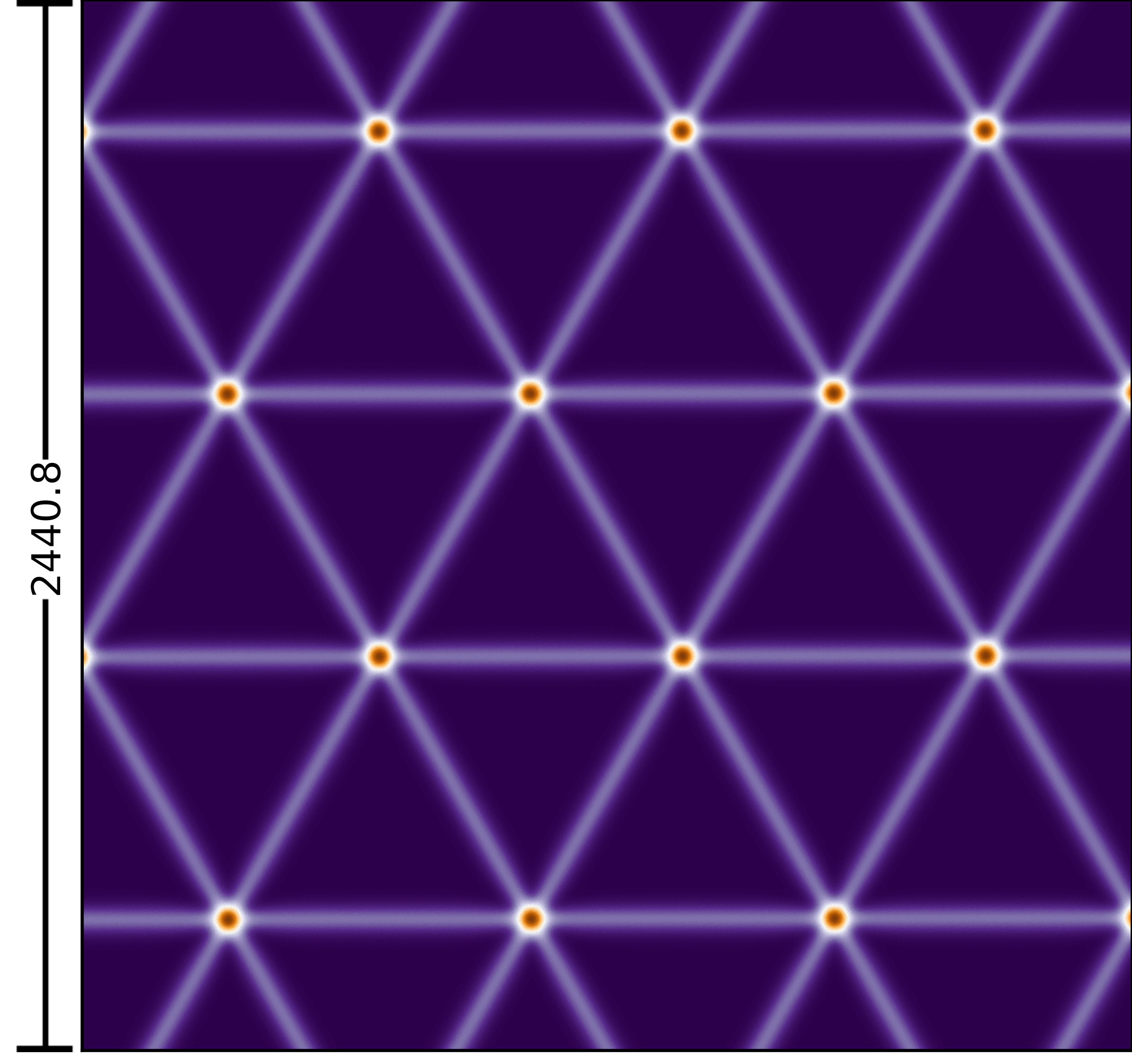} \\
        \smallskip
        \includegraphics[width=.95\textwidth]{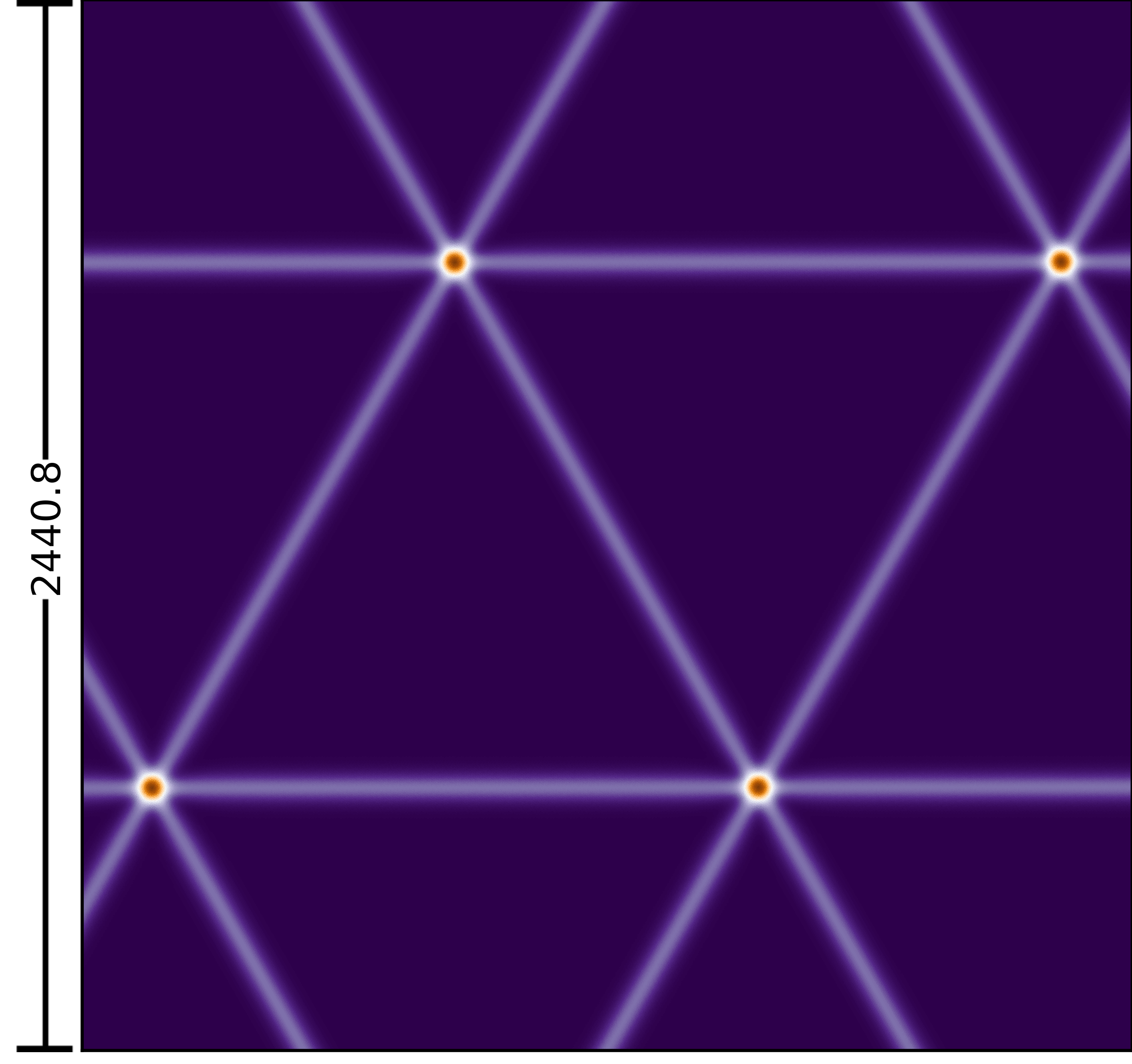}
        }
    \end{minipage}
\caption{Numerical results for twisted bilayer graphene.
 All length units are in \AA. \label{fig:figure10}}
\end{figure}
Resulting features are summarized in Figure~\ref{fig:figure9c}:
\begin{itemize}\setlength\itemsep{0em}
    \item Large triangular domains form where the two layers are in quasi-registry of alternating AB and BA type, minimizing the local interlayer stacking fault energy, with the dominant deformation inside the domains being a local counter-twist of each layer.
    \item These domains are separated by narrow domain walls of shear type, meaning that the interlayer translation (Burgers vector) across the boundary is parallel to the domain wall itself, with three possible orientations, experimentally distinguishable (see Figure~\ref{fig:figure1}).
    \item The domain walls intersect at nodes of AA type (see e.g. Figure~\ref{fig:figure3c}), which due to their nature as maxima of the GSFE are pinned and act as topological defects~\cite{alden2013strain,nonabelian}, and where an additional twist can be observed, dramatically reducing the area of the node.
\end{itemize}

We further present in Figure~\ref{fig:figure10} a numerical study of the moiré pattern using realistic parameters for twisted bilayer graphene as the twist angle decreases from $.8^\circ$ to $.1^\circ$ (see Section~\ref{sec:numerical} for a description of the discretization approach and parameters).
In particular, we note that the domain wall structure, pictured by plotting the order parameter across one of the AB-BA boundaries, converges at angles below $.5^\circ$ to the universal shape and angle-independent width predicted from the analysis in Section~\ref{sec:dw}.

\subsection{Isotropic Strain (Dilation)}
\noindent Consider now the case where the layers are respectively isotropically contracted and dilated.
The deformed lattices and moiré are given by the fundamental matrices:
\[
  A_1 = (1-\varepsilon/2) A, \qquad A_2 = (1+\varepsilon/2) A, \qquad \text{such that} \quad \mathcal{A}_\mathcal{M} = \frac{1-(\varepsilon/2)^2}{\varepsilon} A.
\]
\begin{figure}[h!]
    \centering
    \subcaptionbox{Unrelaxed configuration\label{fig:figure11a}}[.49\textwidth]{\includegraphics[width=.45\textwidth]{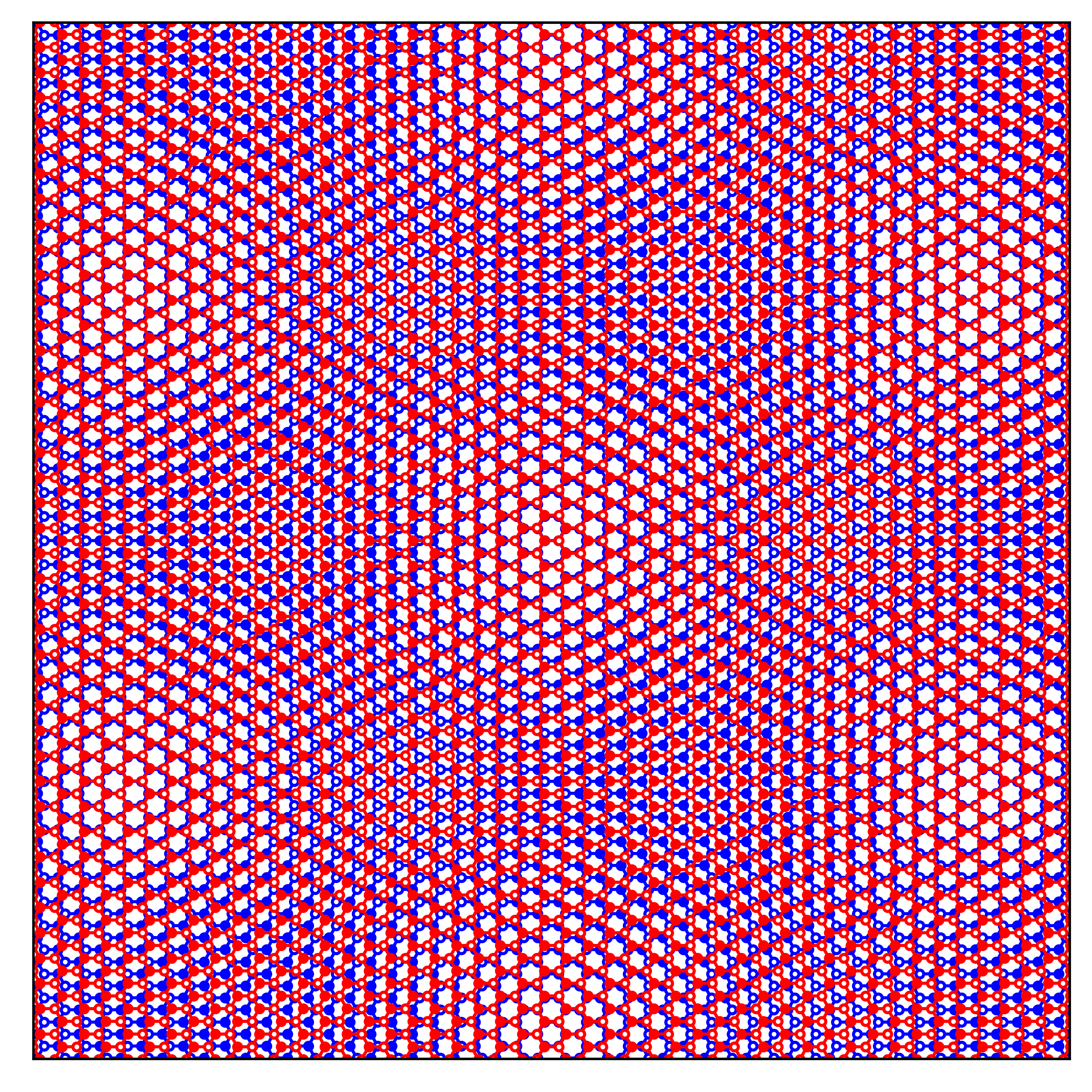}}\\
    \subcaptionbox{Relaxed configuration\label{fig:figure11b}}[.49\textwidth]{\includegraphics[width=.45\textwidth]{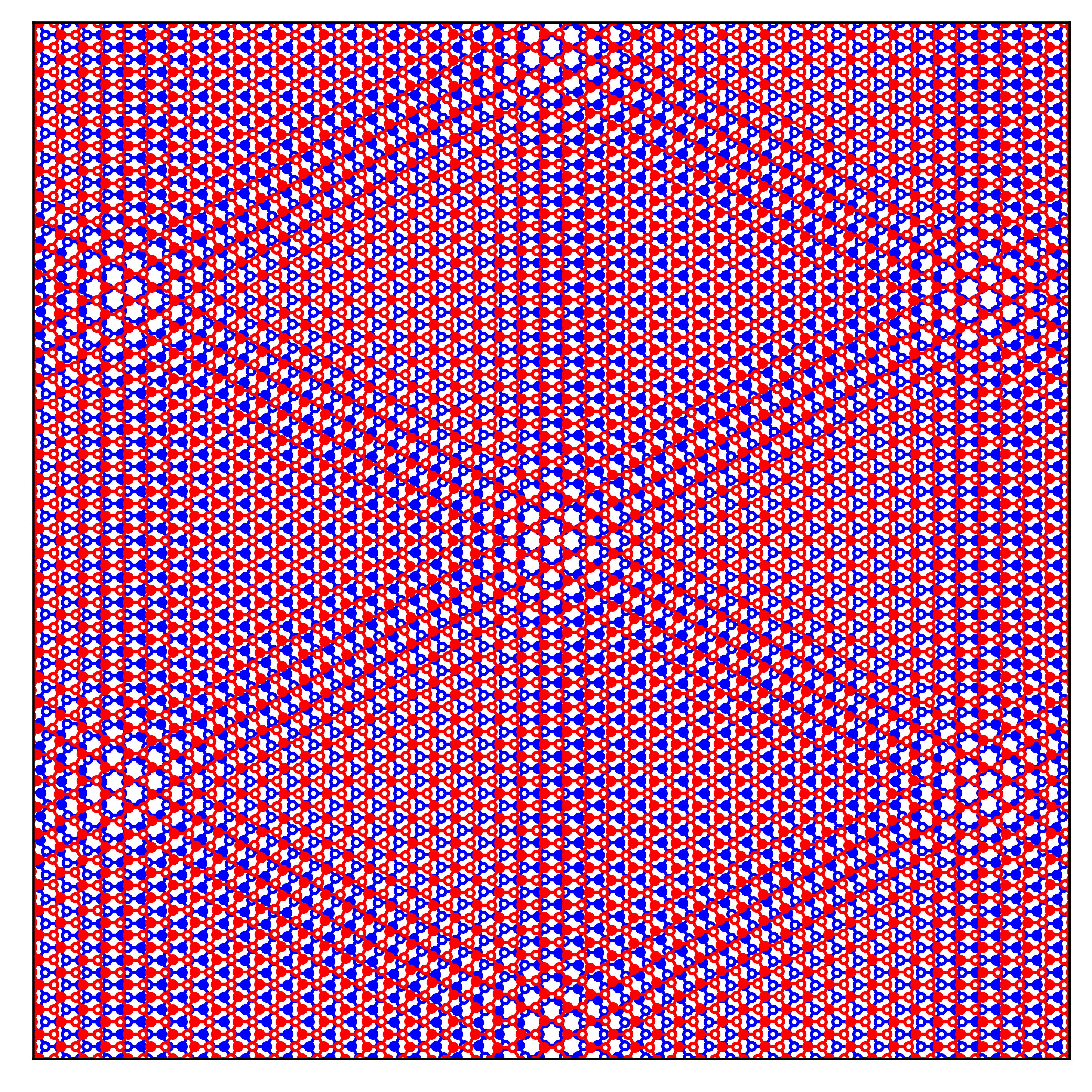}}
    \subcaptionbox{Schematic description of domain formation in dilated bilayer structures\label{fig:figure11c}}[.49\textwidth]
    {\centering \includegraphics[width=.37\textwidth]{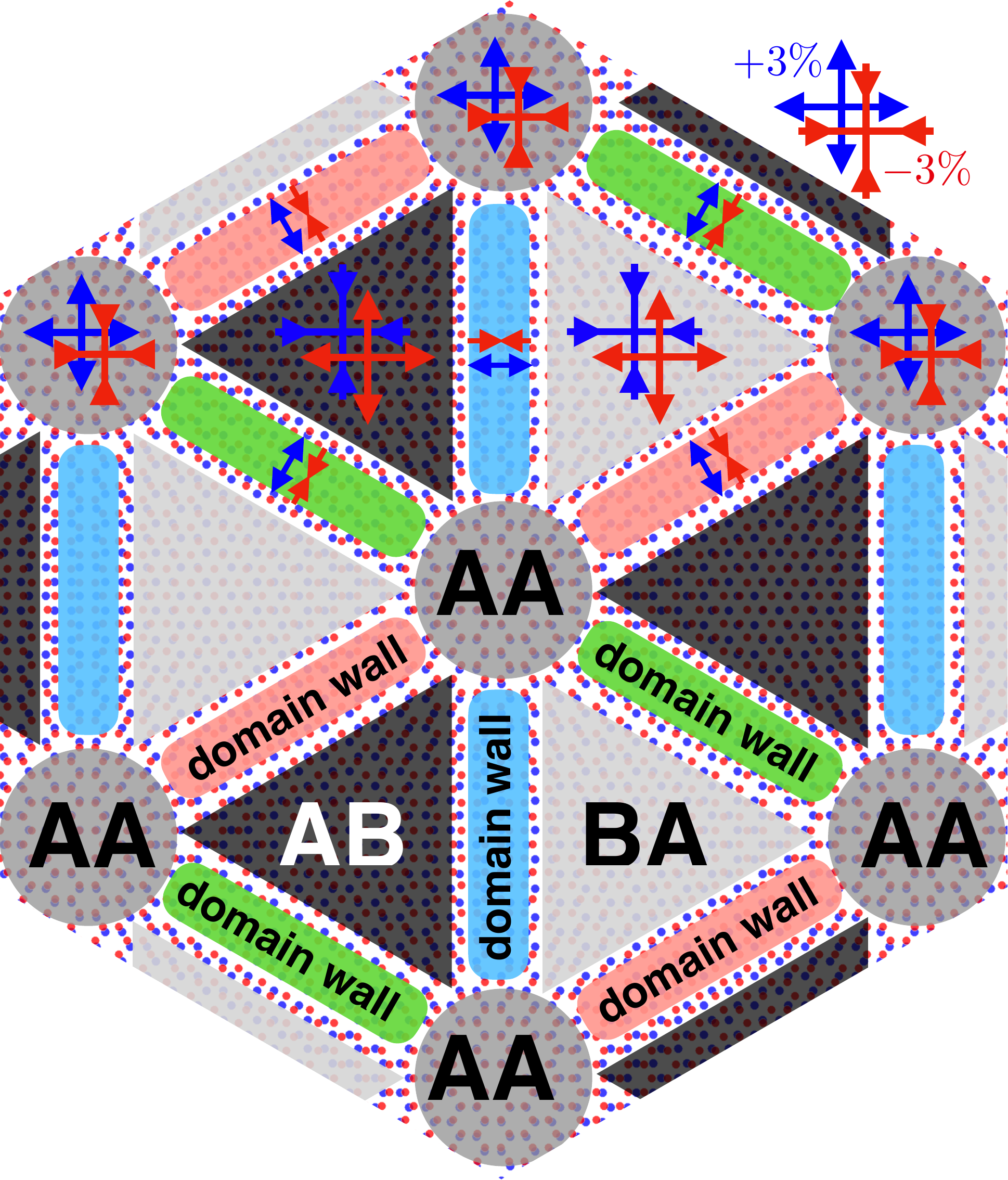}}
    \caption{Isotropic strain moiré ($\varepsilon=6\%$) with misfit energy amplified by a factor of 100.}
    \label{fig:figure11}
\end{figure}\\
\noindent
\noindent The resulting moiré lattice is triangular and aligned with the underlying reference lattice, and scales inversely proportionally to the dilation parameter, $\varepsilon$.
Mathematical analysis is identical to the twisted case above (Theorem~\ref{thm:est}), but the resulting pattern has some differences, as seen on Figure~\ref{fig:figure11c}. In addition to the orientation difference with respect to crystallographic directions (which can be experimentally detected by analyzing the diffraction pattern), the domain walls are tensile boundaries, and thus thicker than in twisted structures as given in the theoretical estimate \eqref{eq:DWcharacteristics}.

\subsection{Pure Shear}

\begin{figure}[b!]
    \centering
    \subcaptionbox{Unrelaxed configuration\label{fig:figure12a}}[.49\textwidth]{\includegraphics[width=.45\textwidth]{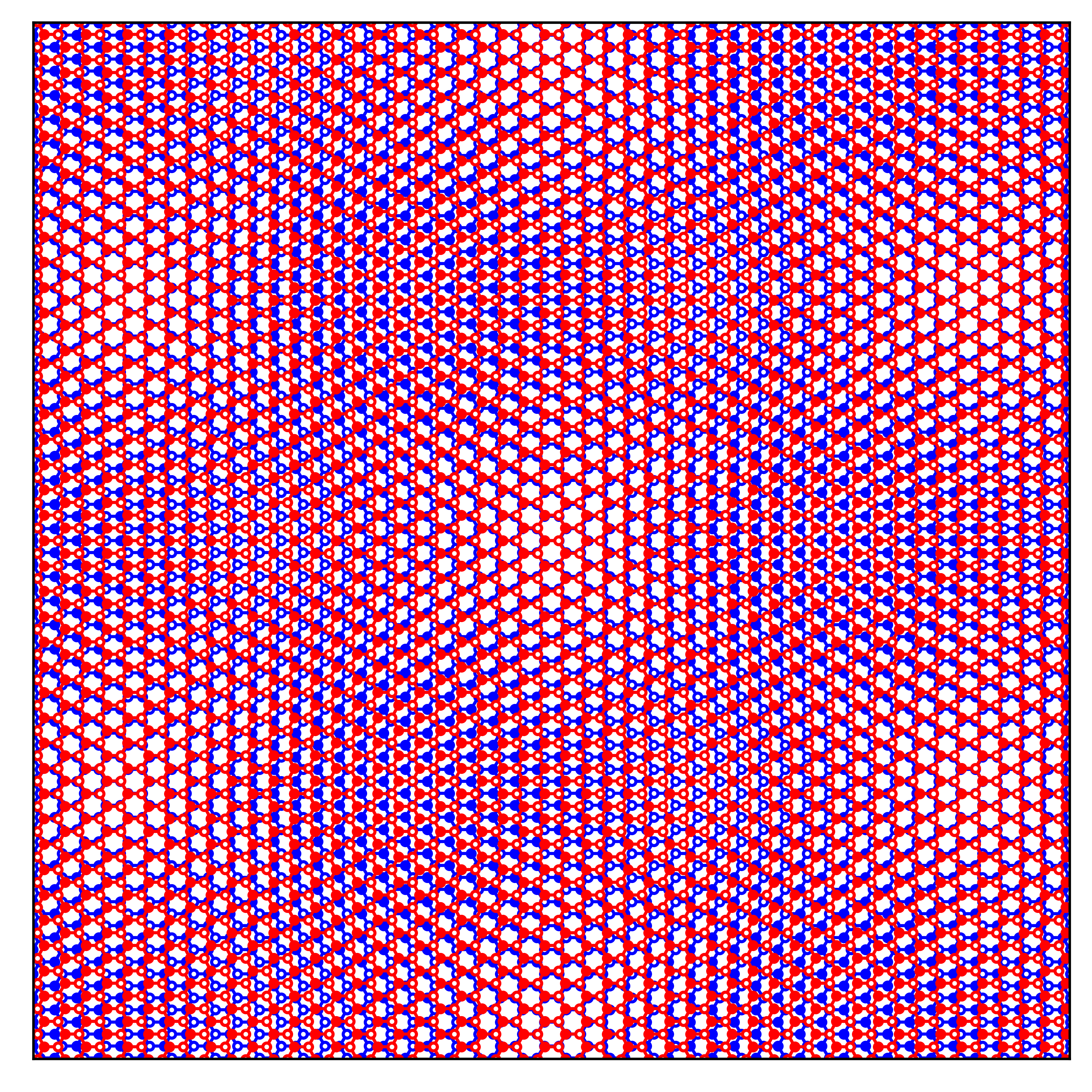}}\\
    \subcaptionbox{Relaxed configuration\label{fig:figure12b}}[.49\textwidth]{\includegraphics[width=.45\textwidth]{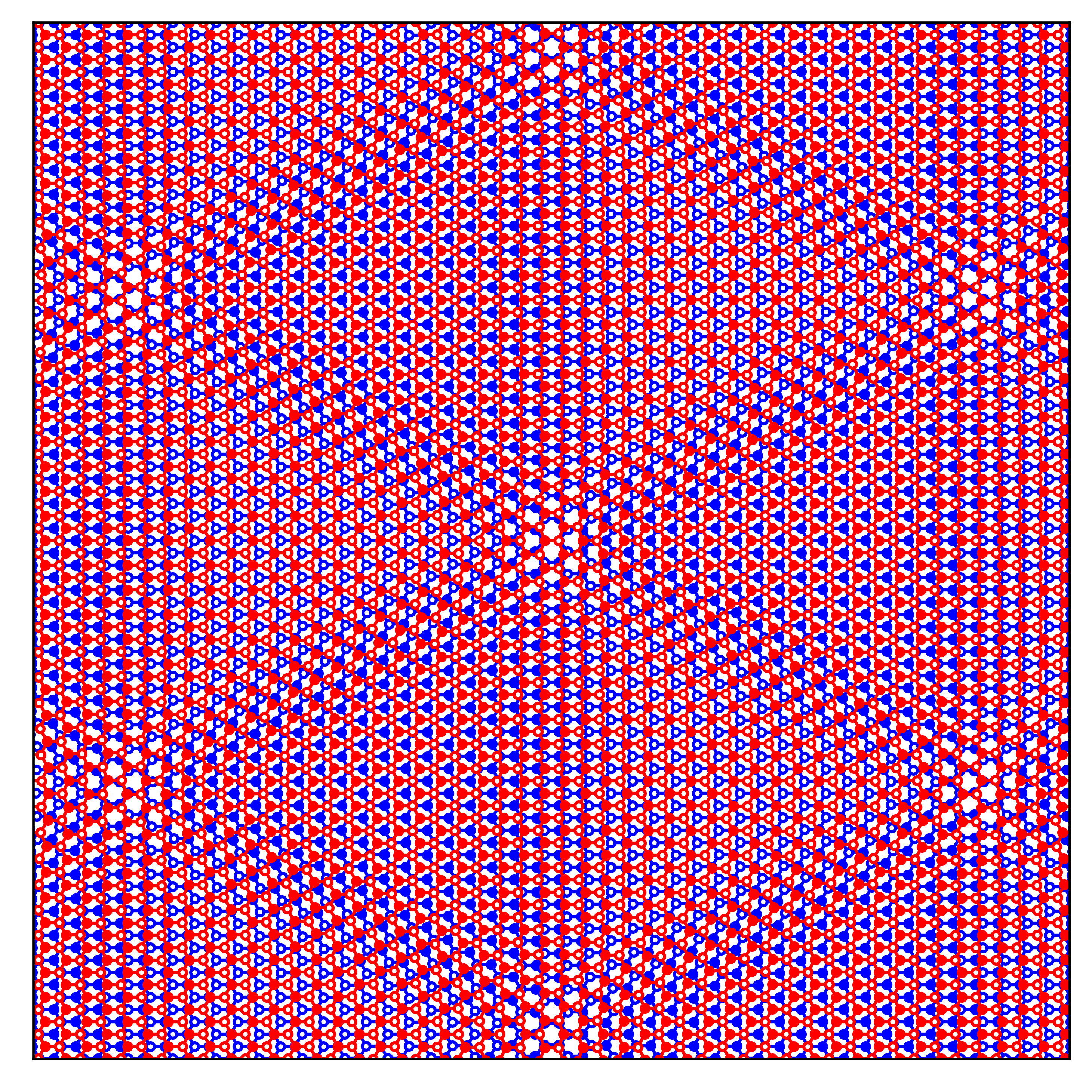}}
    \subcaptionbox{Schematic description of domain formation in the purely sheared bilayer structures\label{fig:figure12c}}[.49\textwidth]
    {\centering \includegraphics[width=.37\textwidth]{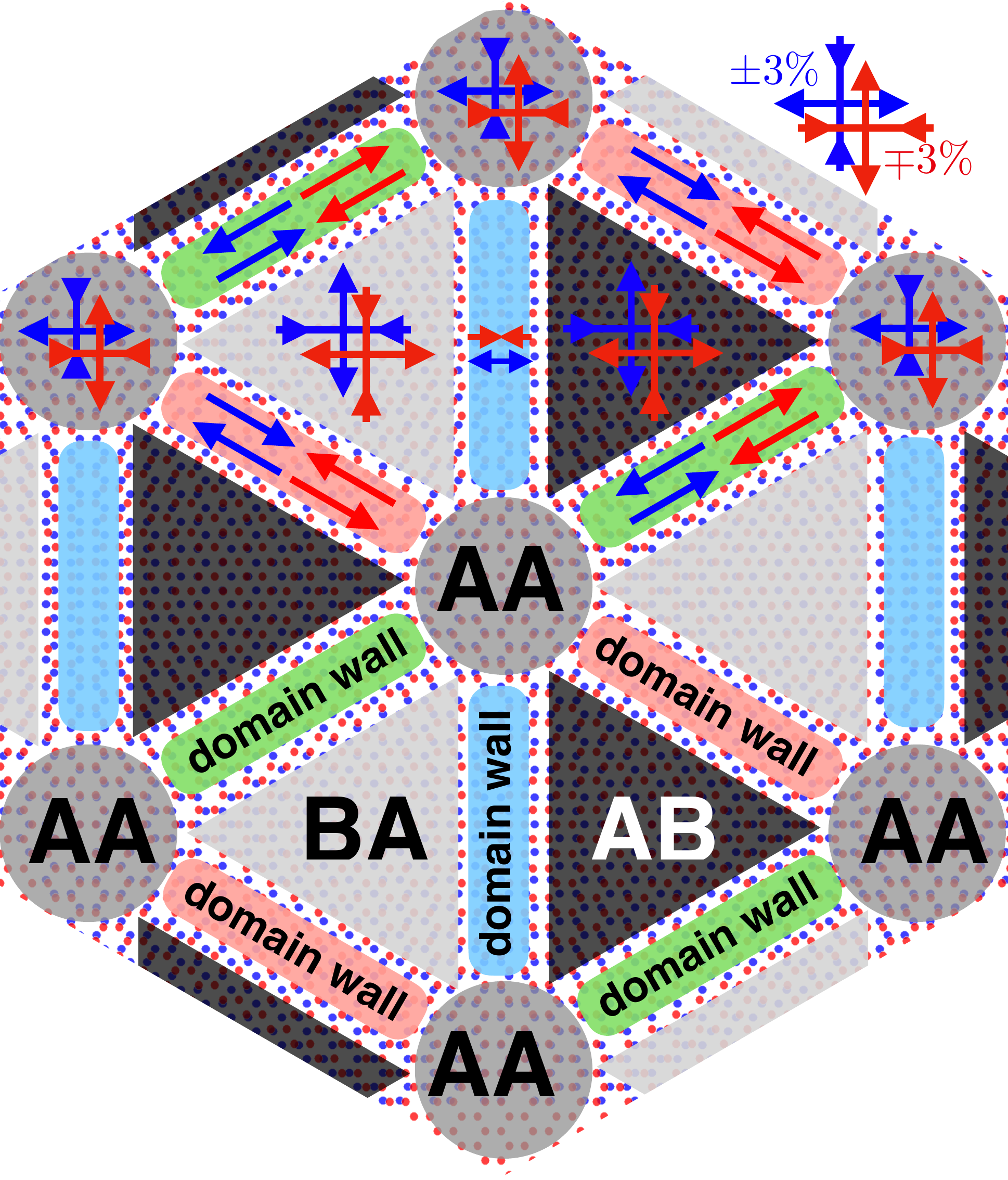}}
    \caption{Purely sheared moiré ($\varepsilon=6\%$) with misfit energy amplified by a factor of 100.}
    \label{fig:figure12}
\end{figure}
\begin{figure}[hp!]
    \begin{minipage}{.37\textwidth}
        \subcaptionbox{
        Relaxed GSFE over pure shear values  $0.0125,$ $0.00625,$ $0.003125,$ and $0.0015625$ (from the top down) scaled and recentered to show the moir\'e scale (left). The \dc{$0.00625$ shear shows the normal to the domain walls (b) and (c) in orange and green.}
        }[\textwidth]
        {\includegraphics[width=.99\textwidth]{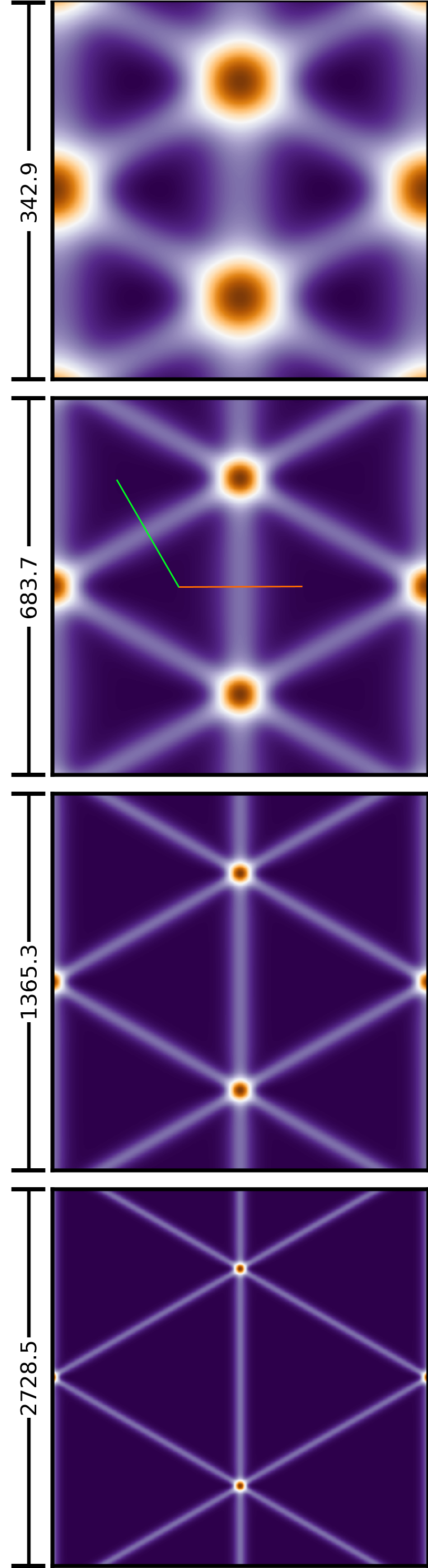}}
    \end{minipage}
    \hspace{.1\textwidth}
    \begin{minipage}{.53\textwidth}
        \subcaptionbox{
        Order parameter $\psi$~\eqref{eq:DWcharacteristics} plotted across a \dc{tensile} AB-BA domain wall for various pure shear values. At~$0.0125$ the plot is cut off beyond the AB and BA points.
            \label{fig:figure13b}
        }[\textwidth]
        {
        \includegraphics[width=\textwidth]{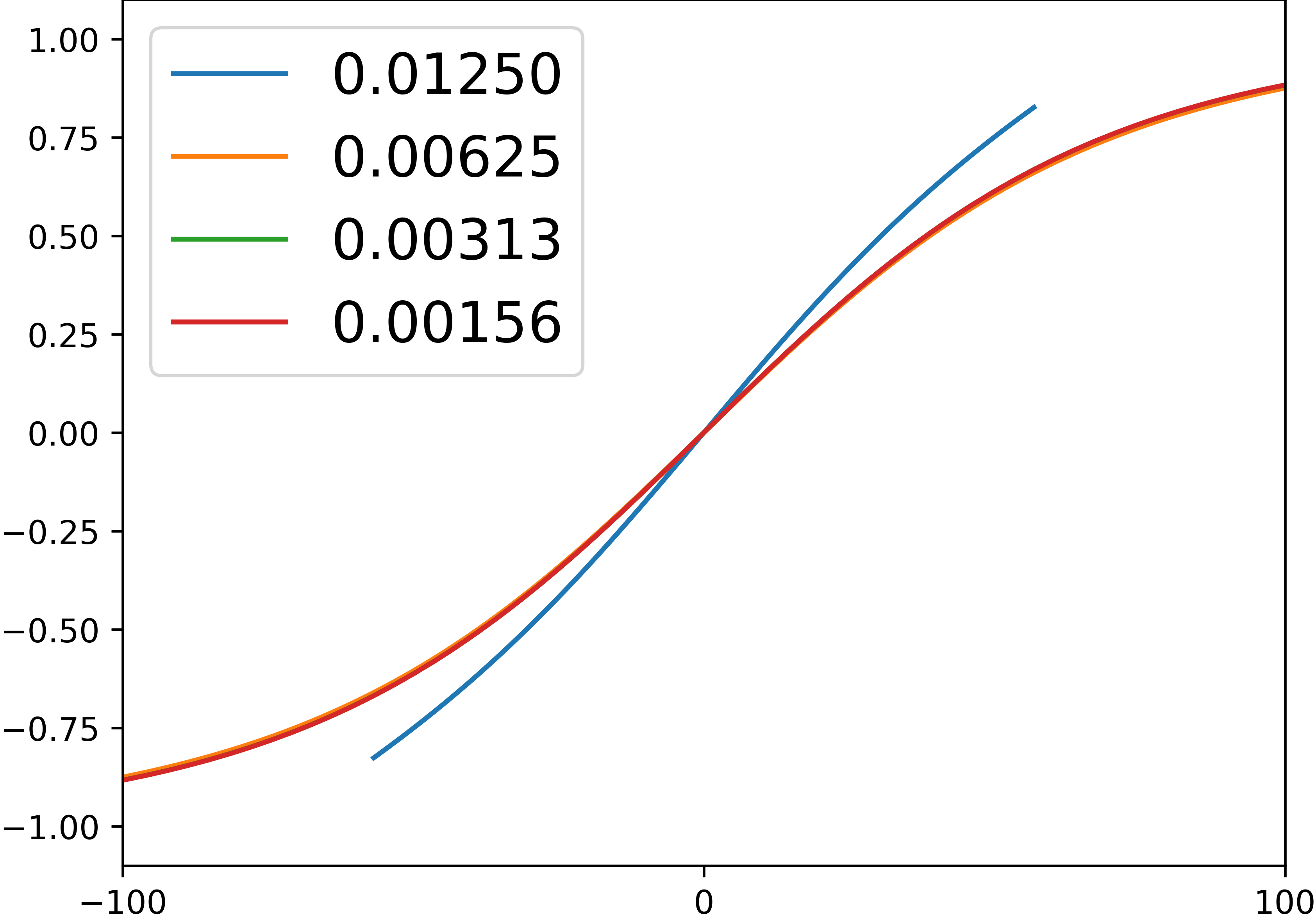}
        }
        \smallskip

        \subcaptionbox{
        \dc{Order parameter $\psi$~\eqref{eq:DWcharacteristics} plotted across a mixed AB-BA domain wall for various pure shear values. At~$0.0125$ the plot is cut off beyond the AB and BA points.
        \label{fig:figure13c}}
        }[\textwidth]
        {
        \includegraphics[width=\textwidth]{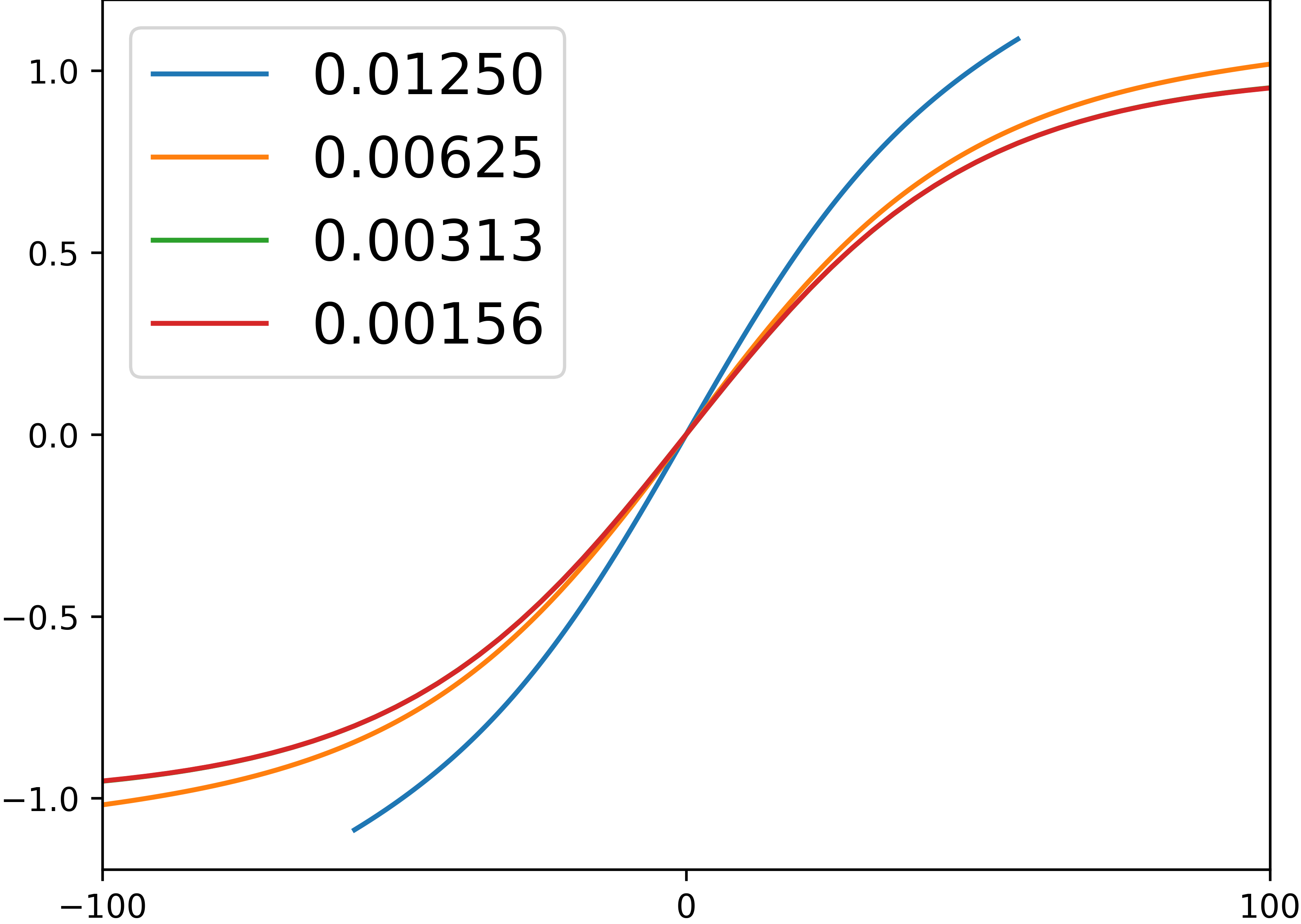}
        }
        \smallskip

        \subcaptionbox{
        Relaxed GSFE with pure shear value of $0.003125$ plotted across the same region \dc{as the plot with shear value $0.0015625$ in (a)} and at a consistent scale.
        }[\textwidth]
        {
        \includegraphics[width=.95\textwidth]{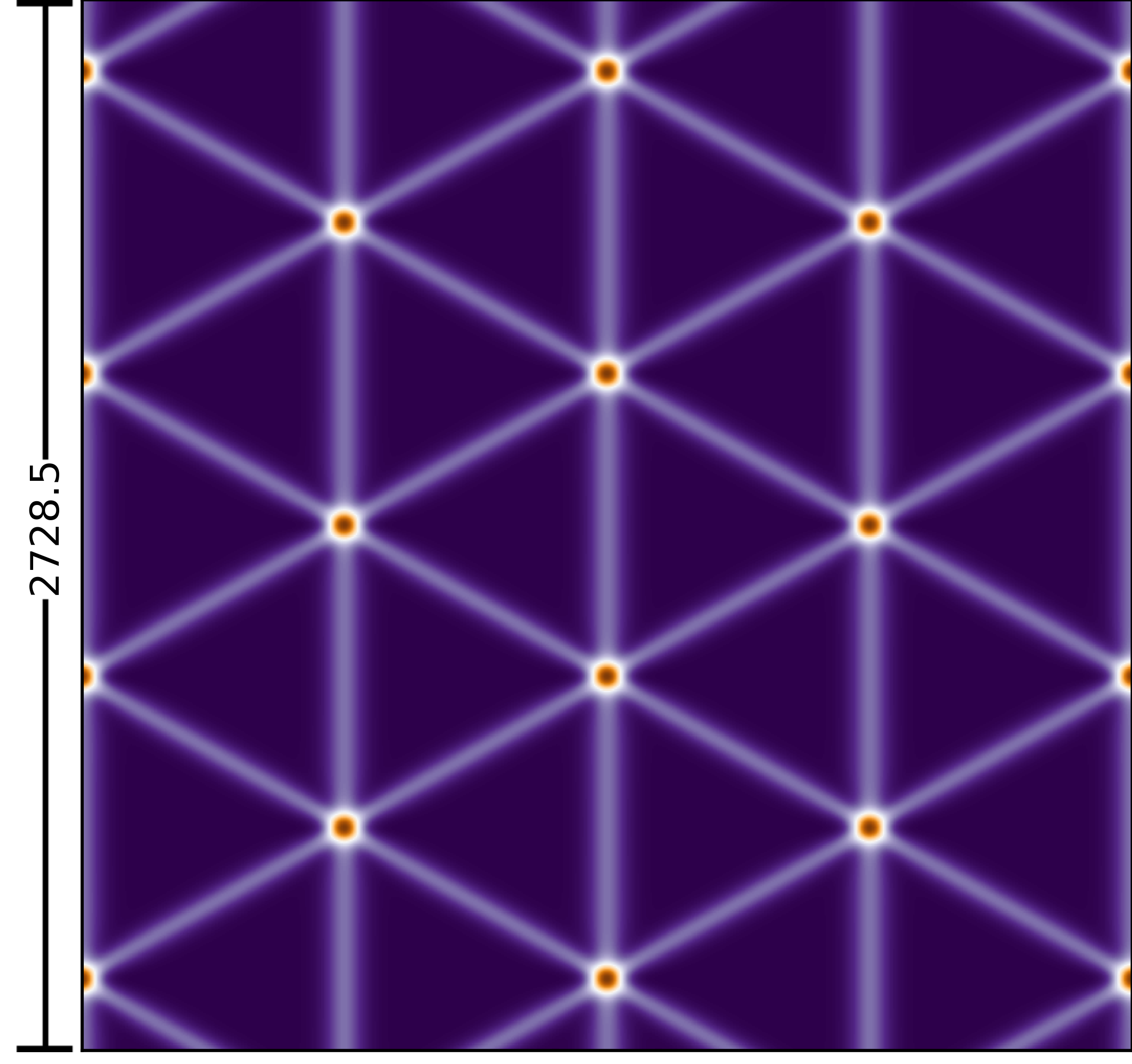} \\
        }
    \end{minipage}
\caption{Numerical results for purely sheared bilayer graphene.  All length units are in \AA. \label{fig:figure13}}
\end{figure}
When the layers are subject to trace-free tensile strain, or pure shear:
\[
  A_1 = \begin{bmatrix} 1-\varepsilon/2 & 0 \\ 0 & 1+\varepsilon/2 \end{bmatrix} A, \
  A_2 = \begin{bmatrix} 1+\varepsilon/2 & 0 \\ 0 & 1-\varepsilon/2 \end{bmatrix} A, \quad \text{s.t.} \
  A_\mathcal{M} = \frac{1-(\varepsilon/2)^2}{\varepsilon} \begin{bmatrix} 1 & 0 \\ 0 & -1 \end{bmatrix} A,
\]
the moiré orientation is obtained by mirror symmetry along the horizontal axis from the underlying crystal lattices, and scales inversely proportionally to the dilation parameter $\varepsilon$.

A defining feature of the relaxation in the pure shear case is that domain walls of a different nature occur throughout the pattern: as seen on Figure~\ref{fig:figure12c}, vertically oriented walls are tensile boundaries while diagonal ones are shear boundaries.
The analysis from Section~\ref{sec:dw} then indicates that the vertical walls will be thicker than the diagonal ones, matching numerical predictions presented in \ml{Figures~\ref{fig:figure13}b and c.}

\begin{remark}
Another important observation is that the AA nodes in pure shear patterns are topologically different from the vertices in twisted or dilated cases, forming so-called anti-vortices~\cite{nonabelian}: indeed, when coloring the domain walls as blue ($B$), green ($G$) or red ($R$) according to the three possible orientations of the interlayer translation across the corresponding boundary (see Figures~\ref{fig:figure1} and~\ref{fig:figure5}), a clockwise loop around vortices encounters domain walls in the order $RGBRGB$ (Fig.~\ref{fig:figure9} and~\ref{fig:figure11}) while a clockwise loop around anti-vortices has the order $RBGRBG$ (Fig.~\ref{fig:figure12}).
\end{remark}

\subsection{Simple Shear}
Finally, consider the case where the layers are subject to horizontal, uniaxial simple shear strain:
\[
  A_1 = \begin{bmatrix} 1 & -\varepsilon/2 \\ 0 & 1 \end{bmatrix} A, \qquad
  A_2 = \begin{bmatrix} 1 &  \varepsilon/2 \\ 0 & 1 \end{bmatrix} A.
\]
This is a particular case where $A_1^{-1} - A_2^{-1}$ is not invertible, indicating that the moiré lattice does not have full rank.
\begin{figure}[b!]
    \centering
    \subcaptionbox{Unrelaxed configuration\label{fig:figure14a}}[.49\textwidth]{\includegraphics[width=.51\textwidth]{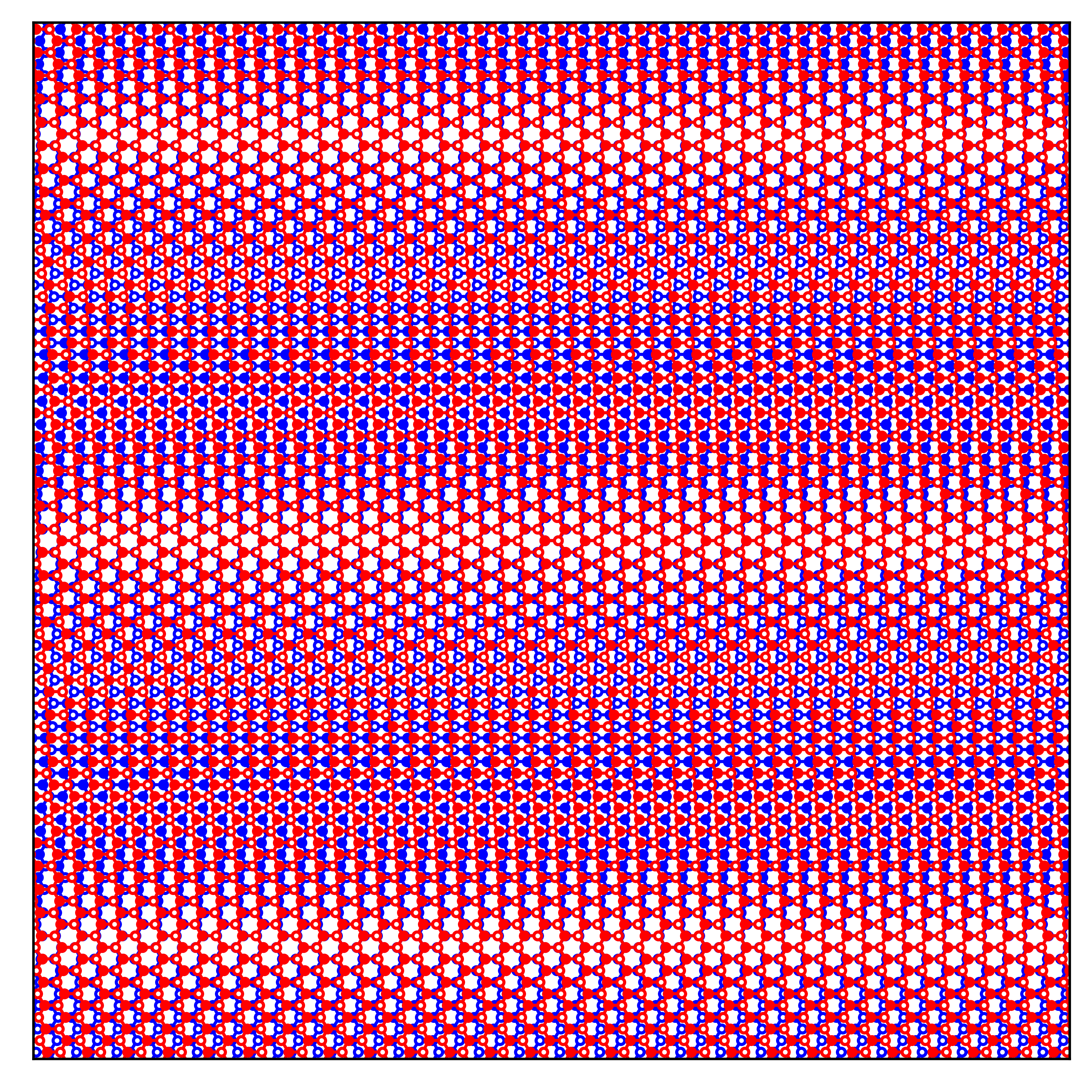}}
    \subcaptionbox{Relaxed configuration\label{fig:figure14b}}[.49\textwidth]{\includegraphics[width=.51\textwidth]{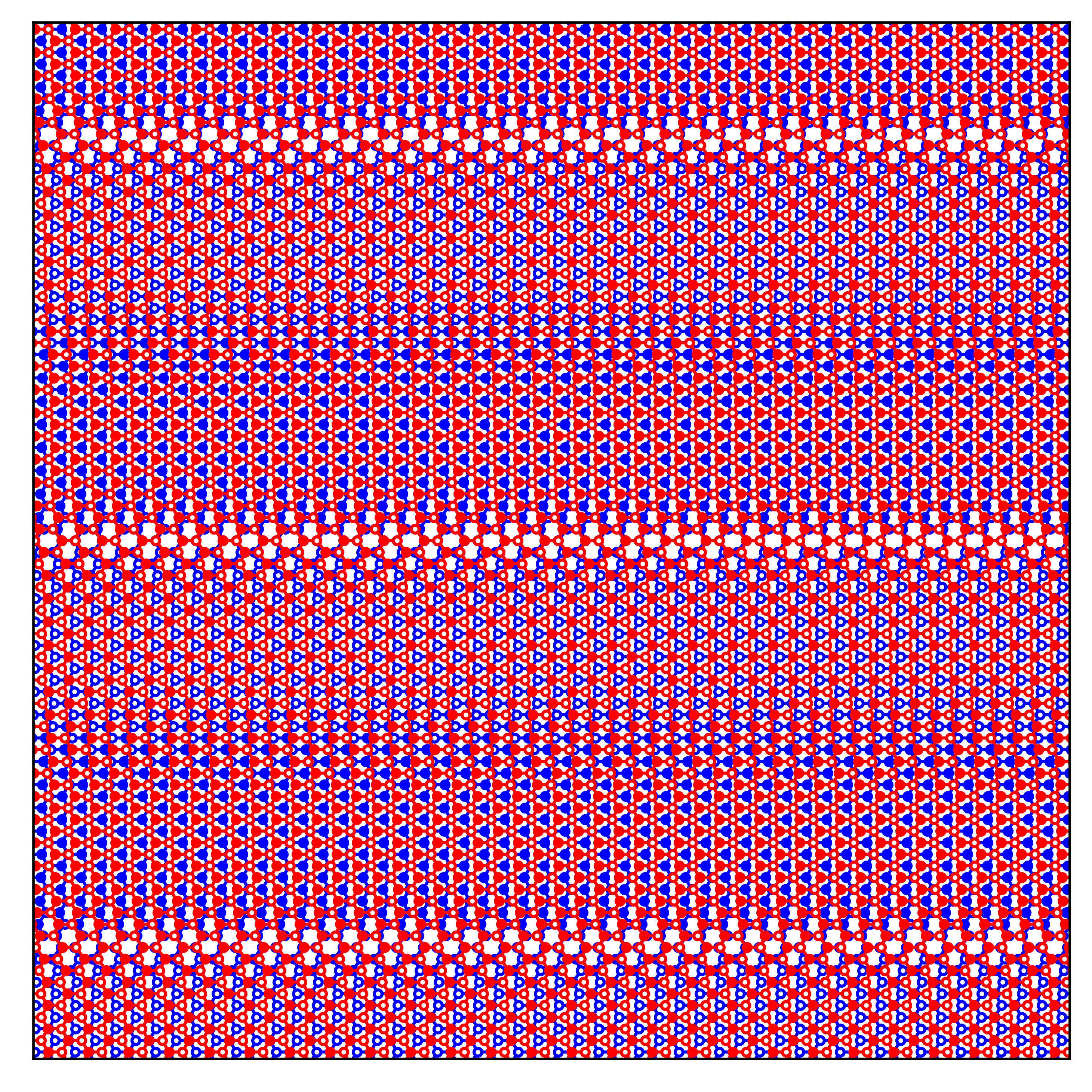}}
    \caption{Relaxation in the simply sheared case ($\varepsilon=10\%$) with misfit energy amplified by a factor of 100.}
    \label{fig:figure14}
\end{figure}

Indeed, as pictured on Figure~\ref{fig:figure8d} we observe a one-dimensional moiré in the vertical direction: generalizing~\eqref{def:moirelattice} as
\[
    A_\mathcal{M} = (A_1^{-1} - A_2^{-1})^\dagger = \begin{bmatrix} 0 & 0 \\ 1/\varepsilon & 0 \end{bmatrix} A
\]
where $(A_1^{-1} - A_2^{-1})^\dagger$ is the Moore-Penrose inverse of $(A_1^{-1} - A_2^{-1}).$

We present on Figure~\ref{fig:figure14} the resulting relaxed pattern. Mathematical analysis in this case is similar to that of the one-dimensional domain walls in Section~\ref{sec:dw}: depending on the initial configuration, a series of horizontal domains alternating between AB and BA registry configurations is created, separated by narrow shear-type boundaries.  \ml{See also \cite{guinea22,Guinea19}}.

Patterns of parallel, elongated domains mirroring~\ref{fig:figure14b} are often observed in experimental images in regions where shear dominates the local strain: see Figures~\ref{fig:figure1} and~\ref{fig:figure7}.

\section{Numerical Method and Study}\label{sec:numerical}

\subsection{Implementation Details}

To demonstrate the above model, we numerically simulate the relaxation of bilayer graphene under each of the four transformations.  Graphene has a triangular multilattice structure with basis:
\[A=\sqrt3 a_0\begin{bmatrix}\frac{\sqrt3}{2}&\frac{\sqrt3}{2} \\ -\frac12 & \frac12\end{bmatrix}\qquad\qquad \text{ where }\quad a_0=1.42\ \text{nm}.\]
The interlayer GSFE potentials $\Phi_1:\Gamma_2\to\R$ and $\Phi_2:\Gamma_1\to\R$ are assumed to have the form \cite{cazeaux2020energy}:
\[\Phi_1(\bgamma_2)=\phi(2\pi A_2^{-1}\bgamma_2),\qquad \Phi_2(\bgamma_1)=\phi(2\pi A_1^{-1}\bgamma_1),\]
where $\phi$ is defined periodically on $[0,2\pi)^2$:
\begin{align*}
    \phi\left(\begin{bmatrix}v \\ w\end{bmatrix}\right) := c_0&+ c_1[\cos(v)+\cos(w)+\cos(v+w)] \\
    &+c_2[\cos(v+2w)+\cos(v-w)+\cos(2v+w)] \\
    &+c_3[\cos(2v)+\cos(2w)+\cos(2v+2w)].
\end{align*}
The Lam\'e parameters $\lambda,\ \mu$ of the graphene layers and the GSFE coefficients $c_{0-3}$ obtained from vdW-DFT calculations~\cite{cazeaux2020energy} are summed up in Table~\ref{tab:GSFEcoeffs}.
\begin{table}[ht]
\centering
\begin{tabular}{|r|r||r|r|r|r|}
\hline
$\lambda$ & $\mu$  & $c_0$ & $c_1$ & $c_2$  & $c_3$  \\ \hline
37,950    & 47,352 & 7.076 & 4.064 & -0.374 & -0.095 \\ \hline 
\end{tabular}
\caption{Elastic moduli and GSFE coefficients for graphene bilayers in units of meV$/$unit cell area (with $c_0$ slightly increased to make the minimum of the GSFE to be $0$).}
\label{tab:GSFEcoeffs}
\end{table}

We directly discretize the minimization problem~\eqref{eq:genPN} to support more general bilayers even though we have shown that for layers with identical elastic moduli, $\bu_1=-\bu_2$.  So, let us define a uniform $N\times N$ grid on the torus $[0,1)^2$ and wave numbers for its plane waves:
\begin{equation}
    \mathcal{G}_N:=\left\{0,\frac{1}{N},\cdots,\frac{N-1}{N}\right\}^2\quad\text{and}\quad \mathcal{G}_N^*:=\left\{0,1,\cdots,N-1\right\}^2.
\end{equation}
We want to compute the unknown displacements $\bu_1^N$ and $\bu_2^N$ for the uniform sampling of the moir\'e unit cell ${A_\mathcal{M}}\Gamma_\mathcal{M}.$ 
We interpolate these nodal values by the Fourier series
\begin{equation}
    \bu_1^N(\bx)=\sum_{\bk\in \mathcal{G}_N^*}\hat{\bu_1}^N_\bk e^{i2\pi \bk\cdot A_\mathcal{M}^{-1}\bx}, \quad\quad
    \bu_2^N(\bx)=\sum_{\bk\in \mathcal{G}_N^*}\hat{\bu_2}^N_\bk e^{i2\pi \bk\cdot A_\mathcal{M}^{-1}\bx},\qquad \bx\in\Gamma_\mathcal{M},
\end{equation}
where
\begin{equation}
    \hat{\bu_1}^N_\bk=\frac 1{N^2}\sum_{\bxi\in \mathcal{G}_N}\bu_1^N(A_\mathcal{M}\bxi) e^{-i2\pi \bk\cdot \bxi}, \quad\quad
    \hat{\bu_2}^N_\bk=\frac 1{N^2}\sum_{\bxi\in \mathcal{G}_N}\bu_2^N(A_\mathcal{M}\bxi) e^{-i2\pi \bk\cdot \bxi},\qquad \bk\in\mathcal{G}_N,
\end{equation}
to obtain extensions of $\bu_1^N$ and $\bu_2^N$ from $A_\mathcal{M}\mathcal{G}_N$ to continuous periodic functions on $\Gamma_\mathcal{M}.$
The elastic energy can then be computed exactly and the misfit energy \eqref{eq:misfitEnergy} can be approximated by uniform quadrature:

\begin{equation*}
\begin{split}
    \mathcal{E}_\mathrm{inter}&(\bu_1-\bu_2)\approx\frac{1}{2N^2}\sum_{\bxi\in\mathcal{G}_N}\Bigg[\phi\left( 2\pi \left(\bxi+A_2^{-1}
    \left(\bu_1^N\left(A_\mathcal{M}\bxi\right)-\bu_2^N\left(A_\mathcal{M}\bxi\right)\right)\right) \right)
    \\&+
    \phi\left( 2\pi \left(\bxi+A_1^{-1} \left(\bu_1^N\left(A_\mathcal{M}\bxi\right)-u_2^N\left(A_\mathcal{M}\bxi\right)\right) \right)\right )\Bigg].
    \end{split}
\end{equation*}


The gradients for these energies can also be explicitly calculated.  We implemented this method in the Julia language with the limited-memory BFGS quasi-Newton algorithm from the Optim.jl library for the numerical minimization of the total energy.

\pc{
\subsection{Numerical Domain Wall Width}
To conclude this section, we present a quantitative investigation of domain wall widths using numerical simulations.
As is common in the literature (see e.g. ~\cite{alden2013strain}) we propose to use the full width at half maximum (FWHM) of the order parameter $u(y)$ (see Eq.~\eqref{eq:DWcharacteristics}) in the sense
\begin{equation}\label{eq:fwhm}
    L = \vert y_+ - y_- \vert \qquad \text{where} \qquad u(y_\pm) = \pm \frac 12,
\end{equation}
where $y$ is the distance to the interface between domains.
Values corresponding to three domain wall types are reported in Table~\ref{tab:table2}, with
\begin{itemize}
    \item a shear-type domain wall observed in a twist moiré, see Fig.~\ref{fig:figure10b};
    \item a tensile-type domain wall observed in a pure shear moiré, e.g., the vertical interface intersected by the red path on Fig.~\ref{fig:figure13};
    \item a domain wall observed in a pure shear moiré, e.g., the oblique interface intersected by the green path on Fig.~\ref{fig:figure13}.
    Such oblique walls are neither of pure shear or tensile type, but have an angle between the interlayer translation direction and domain wall normal equal to $\theta_0 + \phi = \pi/3$ as seen on Fig.~\ref{fig:figure12c}.
\end{itemize}
\begin{table}[h]
    \centering
    \begin{tabular}{|c||c|c|c|c|c|} \hline
         Shear domain wall,         & Twist angle $\theta$          & 0.8º   & 0.4º   & 0.2º   & 0.1º                       \\ \cline{2-6}
         $\theta_0+\phi=0$          & FHWM $L_\perp$ & 37.7         & 47.4   & 48.7   & 48.7                                \\ \hline \hline
                                    & Shear strength $\varepsilon$  & 0.0125 & 0.0063 & 0.0031 & 0.0016                     \\ \cline{2-6}
         Tensile domain wall,       & FHWM $L_\parallel$            & 61.0   & 77.6   & 76.4   & 76.4                       \\
         $\theta_0+\phi=\pi/2$      & Ratio $L_\parallel/L_\perp^0$ & 1.25   & 1.59   & 1.57   & 1.57                       \\  \cline{2-6}
                                    & $l_\parallel/l_\perp$ from ~\eqref{eq:DWcharacteristics} & \multicolumn{4}{c|}{1.571} \\ \hline \hline
                                    & Shear strength $\varepsilon$  & 0.0125 & 0.0063 & 0.0031 & 0.0016                     \\ \cline{2-6}
         Mixed domain wall          & FHWM $L_{\pi/3}$              & 38.1   & 51.5   & 55.2   & 55.1                       \\
         $\theta_0+\phi=\pi/3$      & Ratio $L_{\pi/3}/L_\perp^0$   & 0.78   & 1.06   & 1.13   & 1.13                       \\ \cline{2-6}
                                    & $l_{\pi/3}/l_\perp$ from ~\eqref{eq:DWcharacteristics}  &\multicolumn{4}{c|}{1.169}   \\ \hline
    \end{tabular}
    \caption{Values of the domain wall full width at half maximum (FWHM)~\eqref{eq:fwhm} obtained by numerical computations as in Fig.~\ref{fig:figure10b} and Fig.~\ref{fig:figure13b},~\ref{fig:figure13c}, and numerical comparison of the width ratios between domain walls of different type to the theoretically predicted ratio from Section~\ref{sec:dw} using as limit width $L^0_\perp$ of the shear-type domain wall computed numerically for the smallest twist angle.}
    \label{tab:table2}
\end{table}

Note that the FWHM $L$ of a given domain wall, in the limit of infinite moiré domain size, should be proportional to the characteristic width $l$ given in~\ref{eq:DWcharacteristics}, but the precise ratio between the two depends on the exact GSFE profile: see Prop.~\ref{prop:dw}.
Hence, for a given GSFE profile but different moiré and/or wall types, we can test our domain wall model from Section~\ref{sec:dw} using these numerical simulations, in particular by comparing the numerically obtained ratio $L / L_\perp$ to the theoretical ratio $l_\phi / l_\perp = \sqrt{1 + \frac{\lambda+\mu}{\mu}\cos(\theta_0+\phi)}$.

We conclude that in the case of shear-type and tensile-type domain walls, the theory and numerical computations seem in perfect agreement; however for the oblique domain wall with $\theta_0+\phi = \pi/3$, there is qualitative agreement but a slight numerical deviation. This is due to one of the assumptions of the model from Section~\ref{sec:dw} holding only approximately, namely the actual displacements $\bu(y)$ are not exactly aligned with the interlayer translation $\Delta \bb$.
}

\end{document}